\documentclass[twoside,fleqn,3p,twocolumn,times,longtitle]{elsarticle}

\usepackage{graphicx}
\usepackage{epsfig}
\usepackage{setspace}

\usepackage{lineno}  

\input babarsym.tex

\newcommand{\BABARPubNumber}  {12/031}

\newcommand{\etal}{\mbox{\textsl{et al.}}\xspace}

\newcommand{\vs}{\mbox{\textsl{vs.}}\xspace}

\def\beq{\begin{equation}}
\def\eeq{\end{equation}}

\def\beqa{\begin{eqnarray}}
\def\eeqa{\end{eqnarray}}

\def\lumi{\lum}

\def\sigmavis{\ensuremath{\sigma_{\rm vis}}}

\def\eeMode{\ensuremath{\epem\to\epem}}
\def\mumuMode{\ensuremath{\epem\to\mu^+\mu^-}}
\def\gagaMode{\ensuremath{\epem\to\gaga}}
\def\tautauMode{\ensuremath{\epem\to\tau^+\tau^-}}
\def\thBha{\ensuremath{\theta^{\rm Lab}_{\rm max}}}

\def\off{{\rm off}}

\begin{document}

\begin{frontmatter}

\title{
{\small
\begin{flushleft}
\babar-PUB-\BABARPubNumber\\
SLAC-PUB-15347\\
arXiv:1301.2703 [hep-ex]\\
\end{flushleft}
}
\vspace*{1.5cm}
%
%
Time-Integrated Luminosity Recorded by the \babar\ 
Detector at the \pep2\ \epem\ Collider}

%
\author[Annecy]{J.~P.~Lees}
\author[Annecy]{V.~Poireau}
\author[Annecy]{V.~Tisserand}
\address[Annecy]{Laboratoire d'Annecy-le-Vieux de Physique des Particules (LAPP), Universit\'e de Savoie, CNRS/IN2P3,  F-74941 Annecy-Le-Vieux, France}
\author[Barcelona]{E.~Grauges}
\address[Barcelona]{Universitat de Barcelona, Facultat de Fisica, Departament ECM, E-08028 Barcelona, Spain }
\author[Bari]{A.~Palano$^{ab,}$ }
\address[Bari]{INFN Sezione di Bari$^{a}$; Dipartimento di Fisica, Universit\`a di Bari$^{b}$, I-70126 Bari, Italy }
\author[Bergen]{G.~Eigen}
\author[Bergen]{B.~Stugu}
\address[Bergen]{University of Bergen, Institute of Physics, N-5007 Bergen, Norway }
\author[LBL]{D.~N.~Brown}
\author[LBL]{L.~T.~Kerth}
\author[LBL]{Yu.~G.~Kolomensky}
\author[LBL]{G.~Lynch}
\address[LBL]{Lawrence Berkeley National Laboratory and University of California, Berkeley, California 94720, USA }
\author[Bochum]{H.~Koch}
\author[Bochum]{T.~Schroeder}
\address[Bochum]{Ruhr Universit\"at Bochum, Institut f\"ur Experimentalphysik 1, D-44780 Bochum, Germany }
\author[UBC]{D.~J.~Asgeirsson}
\author[UBC]{C.~Hearty}
\author[UBC]{T.~S.~Mattison}
\author[UBC]{J.~A.~McKenna}
\author[UBC]{R.~Y.~So}
\address[UBC]{University of British Columbia, Vancouver, British Columbia, Canada V6T 1Z1 }
\author[Brunel]{A.~Khan}
\address[Brunel]{Brunel University, Uxbridge, Middlesex UB8 3PH, United Kingdom }
\author[Budker]{V.~E.~Blinov}
\author[Budker]{A.~R.~Buzykaev}
\author[Budker]{V.~P.~Druzhinin}
\author[Budker]{V.~B.~Golubev}
\author[Budker]{E.~A.~Kravchenko}
\author[Budker]{A.~P.~Onuchin}
\author[Budker]{S.~I.~Serednyakov}
\author[Budker]{Yu.~I.~Skovpen}
\author[Budker]{E.~P.~Solodov}
\author[Budker]{K.~Yu.~Todyshev}
\author[Budker]{A.~N.~Yushkov}
\address[Budker]{Budker Institute of Nuclear Physics SB RAS, Novosibirsk 630090, Russia }
\author[Irvine]{D.~Kirkby}
\author[Irvine]{A.~J.~Lankford}
\author[Irvine]{M.~Mandelkern}
\address[Irvine]{University of California at Irvine, Irvine, California 92697, USA }
\author[Riverside]{B.~Dey}
\author[Riverside]{J.~W.~Gary}
\author[Riverside]{O.~Long}
\author[Riverside]{G.~M.~Vitug}
\address[Riverside]{University of California at Riverside, Riverside, California 92521, USA }
\author[SB]{C.~Campagnari}
\author[SB]{M.~Franco Sevilla}
\author[SB]{T.~M.~Hong}
\author[SB]{D.~Kovalskyi}
\author[SB]{J.~D.~Richman}
\author[SB]{C.~A.~West}
\address[SB]{University of California at Santa Barbara, Santa Barbara, California 93106, USA }
\author[SC]{A.~M.~Eisner}
\author[SC]{W.~S.~Lockman}
\author[SC]{A.~J.~Martinez}
\author[SC]{B.~A.~Schumm}
\author[SC]{A.~Seiden}
\address[SC]{University of California at Santa Cruz, Institute for Particle Physics, Santa Cruz, California 95064, USA }
\author[Caltech]{D.~S.~Chao}
\author[Caltech]{C.~H.~Cheng}
\author[Caltech]{B.~Echenard}
\author[Caltech]{K.~T.~Flood}
\author[Caltech]{D.~G.~Hitlin}
\author[Caltech]{P.~Ongmongkolkul}
\author[Caltech]{F.~C.~Porter}
\author[Caltech]{A.~Y.~Rakitin}
\address[Caltech]{California Institute of Technology, Pasadena, California 91125, USA }
\author[Cincinnati]{R.~Andreassen}
\author[Cincinnati]{Z.~Huard}
\author[Cincinnati]{B.~T.~Meadows}
\author[Cincinnati]{M.~D.~Sokoloff}
\author[Cincinnati]{L.~Sun}
\address[Cincinnati]{University of Cincinnati, Cincinnati, Ohio 45221, USA }
\author[Colorado1]{P.~C.~Bloom}
\author[Colorado1]{W.~T.~Ford}
\author[Colorado1]{A.~Gaz}
\author[Colorado1]{U.~Nauenberg}
\author[Colorado1]{J.~G.~Smith}
\author[Colorado1]{S.~R.~Wagner}
\address[Colorado1]{University of Colorado, Boulder, Colorado 80309, USA }
\author[CSU]{R.~Ayad\fnref{Tabuk}}
\fntext[Tabuk]{Now at the University of Tabuk, Tabuk 71491, Saudi Arabia}
\author[CSU]{W.~H.~Toki}
\address[CSU]{Colorado State University, Fort Collins, Colorado 80523, USA }
\author[Dortmund]{B.~Spaan}
\address[Dortmund]{Technische Universit\"at Dortmund, Fakult\"at Physik, D-44221 Dortmund, Germany }
\author[Dresden]{K.~R.~Schubert}
\author[Dresden]{R.~Schwierz}
\address[Dresden]{Technische Universit\"at Dresden, Institut f\"ur Kern- und Teilchenphysik, D-01062 Dresden, Germany }
\author[EPoly]{D.~Bernard}
\author[EPoly]{M.~Verderi}
\address[EPoly]{Laboratoire Leprince-Ringuet, Ecole Polytechnique, CNRS/IN2P3, F-91128 Palaiseau, France }
\author[Edinburgh]{P.~J.~Clark}
\author[Edinburgh]{S.~Playfer}
\address[Edinburgh]{University of Edinburgh, Edinburgh EH9 3JZ, United Kingdom }
\author[Ferrara]{D.~Bettoni$^{a,}$ }
\author[Ferrara]{C.~Bozzi$^{a,}$ }
\author[Ferrara]{R.~Calabrese$^{ab,}$ }
\author[Ferrara]{G.~Cibinetto$^{ab,}$ }
\author[Ferrara]{E.~Fioravanti$^{ab,}$}
\author[Ferrara]{I.~Garzia$^{ab,}$}
\author[Ferrara]{E.~Luppi$^{ab,}$ }
\author[Ferrara]{L.~Piemontese$^{a,}$ }
\author[Ferrara]{V.~Santoro$^{a,}$}
\address[Ferrara]{INFN Sezione di Ferrara$^{a}$; Dipartimento di Fisica, Universit\`a di Ferrara$^{b}$, I-44100 Ferrara, Italy }
\author[Frascati]{R.~Baldini-Ferroli}
\author[Frascati]{A.~Calcaterra}
\author[Frascati]{R.~de~Sangro}
\author[Frascati]{G.~Finocchiaro}
\author[Frascati]{P.~Patteri}
\author[Frascati]{I.~M.~Peruzzi \fnref{Perugia-alt}}
\fntext[Perugia-alt]{Also with Universit\`a di Perugia, Dipartimento di Fisica, Perugia, Italy }
\author[Frascati]{M.~Piccolo}
\author[Frascati]{M.~Rama}
\author[Frascati]{A.~Zallo}
\address[Frascati]{INFN Laboratori Nazionali di Frascati, I-00044 Frascati, Italy }
\author[Genova]{R.~Contri$^{ab,}$ }
\author[Genova]{E.~Guido$^{ab,}$}
\author[Genova]{M.~Lo~Vetere$^{ab,}$ }
\author[Genova]{M.~R.~Monge$^{ab,}$ }
\author[Genova]{S.~Passaggio$^{a,}$ }
\author[Genova]{C.~Patrignani$^{ab,}$ }
\author[Genova]{E.~Robutti$^{a,}$ }
\address[Genova]{INFN Sezione di Genova$^{a}$; Dipartimento di Fisica, Universit\`a di Genova$^{b}$, I-16146 Genova, Italy  }
\author[ITT]{B.~Bhuyan}
\author[ITT]{V.~Prasad}
\address[ITT]{Indian Institute of Technology Guwahati, Guwahati, Assam, 781 039, India }
\author[Harvard]{M.~Morii}
\address[Harvard]{Harvard University, Cambridge, Massachusetts 02138, USA }
\author[Heidelberg]{A.~Adametz}
\author[Heidelberg]{U.~Uwer}
\address[Heidelberg]{Universit\"at Heidelberg, Physikalisches Institut, Philosophenweg 12, D-69120 Heidelberg, Germany }
\author[Humboldt]{H.~M.~Lacker}
\author[Humboldt]{T.~Lueck}
\address[Humboldt]{Humboldt-Universit\"at zu Berlin, Institut f\"ur Physik, Newtonstr. 15, D-12489 Berlin, Germany }
\author[Imperial]{P.~D.~Dauncey}
\address[Imperial]{Imperial College London, London, SW7 2AZ, United Kingdom }
\author[Iowa]{U.~Mallik}
\address[Iowa]{University of Iowa, Iowa City, Iowa 52242, USA }
\author[IowaState]{C.~Chen}
\author[IowaState]{J.~Cochran}
\author[IowaState]{W.~T.~Meyer}
\author[IowaState]{S.~Prell}
\author[IowaState]{A.~E.~Rubin}
\address[IowaState]{Iowa State University, Ames, Iowa 50011-3160, USA }
\author[Hopkins]{A.~V.~Gritsan}
\address[Hopkins]{Johns Hopkins University, Baltimore, Maryland 21218, USA }
\author[Orsay]{N.~Arnaud}
\author[Orsay]{M.~Davier}
\author[Orsay]{D.~Derkach}
\author[Orsay]{G.~Grosdidier}
\author[Orsay]{F.~Le~Diberder}
\author[Orsay]{A.~M.~Lutz}
\author[Orsay]{B.~Malaescu}
\author[Orsay]{P.~Roudeau}
\author[Orsay]{M.~H.~Schune}
\author[Orsay]{A.~Stocchi}
\author[Orsay]{G.~Wormser}
\address[Orsay]{Laboratoire de l'Acc\'el\'erateur Lin\'eaire, IN2P3/CNRS et Universit\'e Paris-Sud 11, Centre Scientifique d'Orsay, B.~P. 34, F-91898 Orsay Cedex, France }
\author[LLNL]{D.~J.~Lange}
\author[LLNL]{D.~M.~Wright}
\address[LLNL]{Lawrence Livermore National Laboratory, Livermore, California 94550, USA }
\author[Liverpool]{J.~P.~Burke}
\author[Liverpool]{J.~P.~Coleman}
\author[Liverpool]{J.~R.~Fry}
\author[Liverpool]{E.~Gabathuler}
\author[Liverpool]{R.~Gamet}
\author[Liverpool]{D.~E.~Hutchcroft}
\author[Liverpool]{D.~J.~Payne}
\author[Liverpool]{C.~Touramanis}
\address[Liverpool]{University of Liverpool, Liverpool L69 7ZE, United Kingdom }
\author[Mary]{A.~J.~Bevan}
\author[Mary]{F.~Di~Lodovico}
\author[Mary]{R.~Sacco}
\author[Mary]{M.~Sigamani}
\address[Mary]{Queen Mary, University of London, London, E1 4NS, United Kingdom }
\author[Holloway]{G.~Cowan}
\address[Holloway]{University of London, Royal Holloway and Bedford New College, Egham, Surrey TW20 0EX, United Kingdom }
\author[Louisville]{D.~N.~Brown}
\author[Louisville]{C.~L.~Davis}
\address[Louisville]{University of Louisville, Louisville, Kentucky 40292, USA }
\author[Mainz]{A.~G.~Denig}
\author[Mainz]{M.~Fritsch}
\author[Mainz]{W.~Gradl}
\author[Mainz]{K.~Griessinger}
\author[Mainz]{A.~Hafner}
\author[Mainz]{E.~Prencipe}
\address[Mainz]{Johannes Gutenberg-Universit\"at Mainz, Institut f\"ur Kernphysik, D-55099 Mainz, Germany }
\author[Manchester]{R.~J.~Barlow\fnref{uddersfield}}
\fntext[uddersfield]{Now at the University of Huddersfield, Huddersfield HD1 3DH, UK }
\author[Manchester]{G.~D.~Lafferty}
\address[Manchester]{University of Manchester, Manchester M13 9PL, United Kingdom }
\author[UMD]{E.~Behn}
\author[UMD]{R.~Cenci}
\author[UMD]{B.~Hamilton}
\author[UMD]{A.~Jawahery}
\author[UMD]{D.~A.~Roberts}
\address[UMD]{University of Maryland, College Park, Maryland 20742, USA }
\author[UMass]{C.~Dallapiccola}
\address[UMass]{University of Massachusetts, Amherst, Massachusetts 01003, USA }
\author[MIT]{R.~Cowan}
\author[MIT]{D.~Dujmic}
\author[MIT]{G.~Sciolla}
\address[MIT]{Massachusetts Institute of Technology, Laboratory for Nuclear Science, Cambridge, Massachusetts 02139, USA }
\author[McGill]{R.~Cheaib}
\author[McGill]{P.~M.~Patel\fnref{Deceased}}
\fntext[Deceased]{Deceased}
\author[McGill]{S.~H.~Robertson}
\address[McGill]{McGill University, Montr\'eal, Qu\'ebec, Canada H3A 2T8 }
\author[Milano]{P.~Biassoni$^{ab,}$}
\author[Milano]{N.~Neri$^{a,}$}
\author[Milano]{F.~Palombo$^{ab,}$ }
\address[Milano]{INFN Sezione di Milano$^{a}$; Dipartimento di Fisica, Universit\`a di Milano$^{b}$, I-20133 Milano, Italy }
\author[UMiss]{L.~Cremaldi}
\author[UMiss]{R.~Godang\fnref{Alabama}}
\fntext[Alabama]{Now at University of South Alabama, Mobile, Alabama 36688, USA }
\author[UMiss]{R.~Kroeger}
\author[UMiss]{P.~Sonnek}
\author[UMiss]{D.~J.~Summers}
\address[UMiss]{University of Mississippi, University, Mississippi 38677, USA }
\author[Montreal]{X.~Nguyen}
\author[Montreal]{M.~Simard}
\author[Montreal]{P.~Taras}
\address[Montreal]{Universit\'e de Montr\'eal, Physique des Particules, Montr\'eal, Qu\'ebec, Canada H3C 3J7  }
\author[Napoli]{G.~De Nardo$^{ab,}$ }
\author[Napoli]{D.~Monorchio$^{ab,}$ }
\author[Napoli]{G.~Onorato$^{ab,}$ }
\author[Napoli]{C.~Sciacca$^{ab,}$ }
\address[Napoli]{INFN Sezione di Napoli$^{a}$; Dipartimento di Scienze Fisiche, Universit\`a di Napoli Federico II$^{b}$, I-80126 Napoli, Italy }
\author[NIKHEF]{M.~Martinelli}
\author[NIKHEF]{G.~Raven}
\address[NIKHEF]{NIKHEF, National Institute for Nuclear Physics and High Energy Physics, NL-1009 DB Amsterdam, The Netherlands }
\author[NotreDame]{C.~P.~Jessop}
\author[NotreDame]{J.~M.~LoSecco}
\address[NotreDame]{University of Notre Dame, Notre Dame, Indiana 46556, USA }
\author[Ohio]{K.~Honscheid}
\author[Ohio]{R.~Kass}
\address[Ohio]{Ohio State University, Columbus, Ohio 43210, USA }
\author[Oregon]{J.~Brau}
\author[Oregon]{R.~Frey}
\author[Oregon]{N.~B.~Sinev}
\author[Oregon]{D.~Strom}
\author[Oregon]{E.~Torrence}
\address[Oregon]{University of Oregon, Eugene, Oregon 97403, USA }
\author[Padova]{E.~Feltresi$^{ab,}$}
\author[Padova]{N.~Gagliardi$^{ab,}$ }
\author[Padova]{M.~Margoni$^{ab,}$ }
\author[Padova]{M.~Morandin$^{a,}$ }
\author[Padova]{M.~Posocco$^{a,}$ }
\author[Padova]{M.~Rotondo$^{a,}$ }
\author[Padova]{G.~Simi$^{a,}$ }
\author[Padova]{F.~Simonetto$^{ab,}$ }
\author[Padova]{R.~Stroili$^{ab,}$ }
\address[Padova]{INFN Sezione di Padova$^{a}$; Dipartimento di Fisica, Universit\`a di Padova$^{b}$, I-35131 Padova, Italy }
\author[Paris6]{S.~Akar}
\author[Paris6]{E.~Ben-Haim}
\author[Paris6]{M.~Bomben}
\author[Paris6]{G.~R.~Bonneaud}
\author[Paris6]{H.~Briand}
\author[Paris6]{G.~Calderini}
\author[Paris6]{J.~Chauveau}
\author[Paris6]{O.~Hamon}
\author[Paris6]{Ph.~Leruste}
\author[Paris6]{G.~Marchiori}
\author[Paris6]{J.~Ocariz}
\author[Paris6]{S.~Sitt}
\address[Paris6]{Laboratoire de Physique Nucl\'eaire et de Hautes Energies, IN2P3/CNRS, Universit\'e Pierre et Marie Curie-Paris6, Universit\'e Denis Diderot-Paris7, F-75252 Paris, France }
\author[Perugia]{M.~Biasini$^{ab,}$ }
\author[Perugia]{E.~Manoni$^{ab,}$ }
\author[Perugia]{S.~Pacetti$^{ab,}$}
\author[Perugia]{A.~Rossi$^{ab,}$}
\address[Perugia]{INFN Sezione di Perugia$^{a}$; Dipartimento di Fisica, Universit\`a di Perugia$^{b}$, I-06100 Perugia, Italy }
\author[Pisa]{C.~Angelini$^{ab,}$ }
\author[Pisa]{G.~Batignani$^{ab,}$ }
\author[Pisa]{S.~Bettarini$^{ab,}$ }
\author[Pisa]{M.~Carpinelli$^{ab,}$\fnref{Sassari}}
\fntext[Sassari]{Also with Universit\`a di Sassari, Sassari, Italy}
\author[Pisa]{G.~Casarosa$^{ab,}$}
\author[Pisa]{A.~Cervelli$^{ab,}$ }
\author[Pisa]{F.~Forti$^{ab,}$ }
\author[Pisa]{M.~A.~Giorgi$^{ab,}$ }
\author[Pisa]{A.~Lusiani$^{ac,}$ }
\author[Pisa]{B.~Oberhof$^{ab,}$}
\author[Pisa]{E.~Paoloni$^{ab,}$ }
\author[Pisa]{A.~Perez$^{a,}$}
\author[Pisa]{G.~Rizzo$^{ab,}$ }
\author[Pisa]{J.~J.~Walsh$^{a,}$ }
\address[Pisa]{INFN Sezione di Pisa$^{a}$; Dipartimento di Fisica, Universit\`a di Pisa$^{b}$; Scuola Normale Superiore di Pisa$^{c}$, I-56127 Pisa, Italy }
% ...



\begin{abstract}

\begin{doublespace}

We describe a measurement of the time-integrated luminosity
of the data collected by the \babar\ experiment at the \pep2\ 
asymmetric-energy \epem\ collider
at the $\Upsilon(4S)$, $\Upsilon(3S)$, and $\Upsilon(2S)$
resonances and in a continuum region below each resonance.
We measure the time-integrated luminosity 
by counting \eeMode\ and
(for the $\Upsilon(4S)$ only) \mumuMode\ candidate events, 
allowing additional photons in the final state.
We use data-corrected simulation to determine the cross
sections and reconstruction efficiencies for these processes,
as well as the major backgrounds.
Due to the large cross sections of \eeMode\ and
\mumuMode, the statistical uncertainties of the measurement
are substantially smaller than the systematic uncertainties.
The dominant systematic uncertainties are due to observed
differences between data and simulation, as well as 
uncertainties on the cross sections.
For data collected on the $\Upsilon(3S)$ and $\Upsilon(2S)$
resonances, an additional uncertainty arises due to
$\Upsilon\to \epem X$ background. 
For data collected off the $\Upsilon$ resonances, 
we estimate an additional uncertainty due to 
time-dependent efficiency variations, which can affect 
the short off-resonance runs.
The relative uncertainties on the luminosities of the on-resonance 
(off-resonance) samples are 
$0.43\%$ ($0.43\%$) for the $\Upsilon(4S)$,
$0.58\%$ ($0.72\%$) for the $\Upsilon(3S)$,
and $0.68\%$ ($0.88\%$) for the $\Upsilon(2S)$.


\end{doublespace}

\end{abstract}



\end{frontmatter}

\clearpage

\begin{doublespace}


\section{Introduction}
\label{sec:intro}

The \babar\ detector~\cite{ref:babar} operated at the \pep2\ asymmetric-energy $\epem$
collider~\cite{pep-ii} and collected physics data from October 1999 until March 2008.  
Most of the data were collected at an \epem\ center-of-mass (CM) energy
$\sqrt{s}$ corresponding to the mass of the $\Upsilon(4S)$
resonance~\cite{ref:pdg}.  This ``on-resonance'' $\Upsilon(4S)$ sample
contains $(464.8 \pm 2.8) \times 10^6$ $\BB$ events~\cite{McGregor:2008ek} 
and is used for the study of $B$-meson
decays, $CP$ violation, and $\Bz-\Bzb$ mixing.
Data samples collected at the $\Upsilon(3S)$ and $\Upsilon(2S)$
resonances in 2008 are used for bottomonium studies and for dedicated
new-physics searches.
For each $\Upsilon(nS)$ resonance ($n=2,3,4$), an ``off-resonance''
sample was collected for studying continuum $\epem \to q\bar q$
events, where $q$ is a $u$, $d$, $s$, or $c$ quark.  The off-\FourS\
sample has a CM energy about 40~\mev\ below the \FourS\ peak mass,
and the off-\ThreeS\ and off-\TwoS\ samples are 30~\mev\ below the
respective peaks.
All on- and off-resonance samples are used for charm, $\tau$,
two-photon, and QCD physics analyses.

Measurements of production cross sections and 
branching fractions often depend on knowledge of the time-integrated
luminosity $\lumi$ of the collected data sample.  In some cases, the
uncertainty on \lumi\ is one of the major sources
of systematic uncertainty~\cite{ref:lumi-major-error}. 
In addition, in $\Upsilon$-resonance data analyses, background
characteristics or the level of continuum background contamination are
often determined from the off-resonance sample. This 
requires knowledge of the ratio of the integrated luminosities of the
on-resonance and off-resonance samples.

In this article, we describe the final analysis of the integrated
luminosity of the dataset collected by \babar, incorporating the
latest processing and reconstruction of the dataset, improved techniques,
and reduced systematic uncertainties relative to previous 
measurements.
The integrated luminosity is measured with Bhabha (\eeMode) and
dimuon (\mumuMode) events.  These processes have large, well-known
cross sections and simple signatures that are easily identified, thus
ensuring high signal-to-background ratios. We  use diphoton
(\gagaMode) events to estimate some systematic uncertainties and
in the determination of the $\Upsilon(2S, 3S)\to\epem$ background contamination.
We do not use diphoton events to directly
measure the integrated luminosity, due to the
significant uncertainty on the
cross section for this process, as calculated by available
Monte Carlo (MC) generators.

The analysis technique and results are presented here as a resource
for future \babar\ physics publications, as well as future
integrated-luminosity measurements at other \epem\ colliders.

\section{Detector and Dataset}
The \babar\ detector is described in detail in Ref.~\cite{ref:babar},
and only a brief description is given here. 
Charged-particle trajectories are measured with a
five-layer silicon vertex tracker and a 40-layer drift chamber (DCH) in a
nearly uniform 1.5~T magnetic field.  
Charged hadron identification is provided by a Cherenkov detector, and 
photons and electrons are detected
in a CsI(Tl) electromagnetic calorimeter (EMC). 
Muons are identified with resistive plate chambers and limited
streamer tubes inserted between the iron layers of the magnetic-field 
instrumented flux return (IFR).

A two-level trigger system, composed of a hardware (``level-1'') stage and a
subsequent software (``level-3'') stage, is used to decide whether an
event is recorded.
Both trigger levels use information from the DCH and EMC
and employ fast EMC-cluster and track-reconstruction algorithms.
IFR information is also used in level~1.
Events passing the level-1 and level-3 trigger selections are
recorded. After additional prescaling (discussed below), events are
processed by the offline reconstruction, where more
sophisticated algorithms use information from all detector subsystems.
After initial stages of the offline reconstruction, 
an event selection and classification
stage referred to as the offline filter takes place.
Classifications of the level-3 trigger and the offline filter
are used to preselect events for subsequent data analysis.

The integrated luminosity and its uncertainties are determined
separately for several data samples.  The $\Upsilon(4S)$ sample is
divided into six runs, labeled Run~1 through Run~6.  Each run
corresponds to a data-taking period with typical shutdowns of no more
than a few days or weeks.  Shutdown periods between runs are typically
several months long.
For each run there is also an off-resonance sample, collected during
short periods interleaved with on-resonance data-taking periods.
The Run-7 sample contains the $\Upsilon(3S)$ and $\Upsilon(2S)$ data,
as well as the corresponding off-resonance samples.
Run~7 also includes a dataset collected at CM
energies above the \FourS\ resonance, which is not included in this
analysis. 
Table~\ref{tab:periods} lists the data-taking period and $\Upsilon$
resonance for each run.
%

%
\begin{table}[htbp]
\caption{\label{tab:periods} Data-taking period and the resonance
corresponding to the \pep2\ CM energy $\sqrt{s}$ for each of the \babar\
runs. }
\vspace{4mm}
   \centering
   \begin{tabular}{c|c|c}
     \hline\hline
     Resonance & Run & Month/Year \cr
     \hline
     $\Upsilon(4S)$ & Run 1 & 10/1999 $-$ 10/2000  \cr
                    & Run 2 & 02/2001 $-$ 06/2002  \cr
                    & Run 3 & 12/2002 $-$ 06/2003  \cr
                    & Run 4 & 09/2003 $-$ 07/2004  \cr
                    & Run 5 & 04/2005 $-$ 08/2006 \cr
                    & Run 6 & 01/2007 $-$ 09/2007  \cr
     \hline
     $\Upsilon(3S)$ & Run 7 & 12/2007 $-$ 02/2008  \cr
     \hline
     $\Upsilon(2S)$ & Run 7 & 02/2008 $-$ 03/2008  \cr
     \hline\hline
   \end{tabular}
\end{table}

To calculate cross sections and detector efficiencies, we make use of
simulated MC samples. 
The {\tt BHWIDE}~\cite{Jadach:1995nk} MC generator is used to simulate
Bhabha events, and the {\tt KKMC}~\cite{Jadach:1999vf} generator 
with the modifications described in Ref.~\cite{Banerjee:2007is} 
is
used for dimuon events.  We also use {\tt KKMC} to study possible
background from \tautauMode\ events.  The {\tt BABAYAGA} generator
with next-to-leading-order corrections~\cite{Balossini:2006wc} is used
to estimate the Bhabha cross section systematic uncertainty.  The {\tt
EvtGen}~\cite{ref:evtgen} generator is used for studying the
background from $\Upsilon(2S)$ and $\Upsilon(3S)$ decays in Run~7.  We
use the {\tt BKQED}~\cite{ref:bkqed} generator to generate diphoton
events.
Events produced by these MC generators are passed through a full
detector simulation based on {\tt Geant4}~\cite{ref:GEANT} and
are reconstructed and analyzed in the same way as the data.

\section{Analysis Method}
\label{sec:analysys-method}

For Runs~1--6, the integrated luminosity is measured with Bhabha
(\eeMode) and dimuon (\mumuMode) events, which may include any number
of radiated photons in the final state.  For Run~7, \mumuMode\ events
are not used, due to significant uncertainty associated with the 
contribution of the $\Upsilon\to\mu^+\mu^-$ background.

For a particular data sample, the integrated luminosity is measured from
\beq
\lumi = {N_{\rm cand} - N_{\rm bgd} \over \sigmavis},
\label{eq:lum-basic}
\eeq
where $N_{\rm cand}$ is the number of selected signal candidate events,
of which $N_{\rm bgd}$ events are estimated to be background.
The visible cross section $\sigmavis$ is given by
\beq
\sigmavis \equiv \int {d\sigma \over d\Omega} \epsilon(\Omega)\, d\Omega,
\label{eq:sigmavis}
\eeq
where $d\sigma / d\Omega$ is the theoretical differential cross
section and $\epsilon(\Omega)$ the efficiency for reconstructing and
selecting signal events for a given phase-space point $\Omega$.
The methods for obtaining each of these quantities are discussed below.

\subsection{Event Selection}
\label{sec:cuts}

The event-selection criteria are designed to yield samples of
high-purity Bhabha and dimuon events, with two high-momentum
charged-particle tracks in the central part of the detector and
relatively little energy taken up by radiated photons.  
We have chosen the
selection criteria so that systematic uncertainties arising from 
data-MC differences of event distributions are kept to a minimum.
Electron \vs muon identification relies on comparison of the track
momentum with the corresponding energy deposited in the EMC.  Event
selection is performed in two steps: preselection, which takes place
at the level-3 trigger and during offline reconstruction, and is
described in Section~\ref{sec:cuts_pre}; and final event selection,
which is described in Section~\ref{sec:cuts_final}.

As a basic requirement for tracks at both selection steps, 
the point of closest approach of
the track to the incoming \pep2\ beams is required to be less than
1.5~cm in the radial direction ($r$) and less than 10~cm in the beam
direction ($z$).

\subsubsection{Preselection}
\label{sec:cuts_pre}

Tracks used for the level-3 Bhabha event selection must 
have laboratory-frame polar-angle values 
between $0.9\rad$ and $2.5\rad$.
Most Bhabha events are selected by finding two
oppositely charged
tracks with CM momenta above 2.0\gevc, where at least
one of the tracks is associated with an EMC cluster with CM energy of at least
2.5\gev.  
The CM momenta, polar angles, 
and azimuthal
angles of the two tracks are required to satisfy $p_1 + p_2 > 7\gevc$,
$|\theta_1 + \theta_2 - \pi| < 0.5\rad$, and $|\phi_1 - \phi_2 - \pi| <
0.3\rad$. 
To maintain high efficiency, the level-3 Bhabha selection
also accepts events with a single track, provided there is an
EMC energy deposition in the expected location, opposite
the track in the CM frame.
In this case, the requirements on the 
track momentum, the cluster energy, and the polar and azimuthal angles
of the track and cluster are
$p_{\rm track} + E_{\rm cluster} > 6\gev$, 
$|\theta_{\rm track} + \theta_{\rm cluster} - \pi| < 0.2\rad$, 
and $|\phi_{\rm track} + \phi_{\rm cluster} - \pi| < 0.3\rad$,
where these quantities are evaluated in the CM frame.

Bhabha events are recorded not only for luminosity determination, but
also for EMC calibration. The Bhabha cross section increases steeply
with decreasing \epem\ scattering angle. Therefore, a large fraction
of events in regions of high cross section is discarded in order to
reduce the rate of events handled by the data-acquisition system 
without significant detrimental
impact on calibration.  This is achieved by assigning each
trigger-selected Bhabha event to one of seven bins according to
\thBha, the larger of the laboratory-frame polar angles of the two
leptons.  For each bin $i$, only one of every $N_i$ events is logged,
where the ``prescale factor'' $N_i$ increases with \thBha.  This
results in a sawtooth distribution of $\cos\thBha$ that is nonetheless
more uniform  than the original distribution and more suitable for
calibration purposes.  The prescale factor applied to each saved event
is later used to recreate the initial 
$\left|\cos\theta\right|$
spectrum for use in the luminosity determination.

Dimuon events are passed by the level-3 trigger based on
a very loose criterion of a single track with transverse momentum 
$p_T > 0.6\gevc$ (a value further reduced for Run~7) or two tracks,
each having $p_T > 0.25\gevc$. 
This loose selection is possible due to the fact that the 
$\mumuMode$ cross section is much lower than the $\eeMode$ cross section.
At the offline-filter stage, 
dimuon event selection requires two oppositely charged tracks.
The CM momenta of the higher-momentum and lower-momentum tracks
must satisfy $p_1 > 4\gevc$ and 
$p_2 > 2\gevc$, respectively;
the sum of the CM polar angles of the tracks is required to satisfy
$2.8 < \theta_1 + \theta_2 < 3.5\rad$;
and the sum of the CM energies of the EMC clusters 
associated with the two tracks must be
less than 2\gev.

The diphoton level-3 trigger selection requires two EMC clusters.
During Run~1 data collection, the CM energy of each cluster was
required to be at least 0.35 of the \pep2\ CM energy $\sqrt{s}$.
For Runs~2--7, the requirement was decreased to $0.3\sqrt{s}$. 
The sums of the polar and azimuthal angles of the
clusters must satisfy $|\theta_1 + \theta_2 - \pi| < \alpha_0$ and
$||\phi_1 - \phi_2| - \pi| < \alpha_0$ in the CM frame, where $\alpha_0
= 0.5\rad$ for Run~1 and 0.1\rad for Runs~2--7.  The trigger is rejected if
the event has a track with $p_T > 0.25\gevc$.

To facilitate offline checks of simulated trigger efficiency, a
heavily prescaled, unbiased sample of all events satisfying the
level-1 trigger is logged.   For corresponding
checks of the offline-filter stage, a prescaled sample of all
logged events is kept regardless of whether any offline-filter selection
is satisfied.  The use of these ``bypass'' samples is discussed
in Section~\ref{sec:sigma}.

\subsubsection{Final Selection}
\label{sec:cuts_final}

The Bhabha and dimuon event selections for the luminosity analysis
impose additional, tighter final-selection criteria,
relying on event properties obtained with the offline reconstruction.

For Bhabha candidates, 
the CM polar angles of the tracks are required to satisfy 
$\left|\cos\theta\right|<0.70\rad$ for one track and 
$\left|\cos\theta\right|<0.65\rad$ for the other track.
We require
$P_1 > 0.75$ and 
$P_2 > 0.50$, 
where the scaled momentum $P_i \equiv 2 p_i / \sqrt{s}$ is
twice the ratio of the CM momentum $p_i$
of track $i$ to the \pep2\ CM energy $\sqrt{s}$, 
and the index $i=1$ ($i=2$) denotes
the track with the higher (lower) CM momentum.
The acolinearity angle $\alpha$, defined as $180^\circ$ minus the CM
angle between the two tracks, is required to satisfy $\alpha <
30^\circ$.
We attempt to geometrically associate each track with an EMC
cluster and calculate the ratio of the cluster energy to the track
momentum in the laboratory frame. Denoting the higher (lower) ratio
with $(E/p)_H$ ($(E/p)_L$), we require $(E/p)_H > 0.7$ and $(E/p)_L > 0.4$.
If only one track is associated with a cluster, it must
satisfy $(E/p) > 0.7$. Events with no track-cluster association are 
rejected.

For dimuon candidates, we require
$\left|\cos\theta\right|<0.70\rad$ for one track and 
$\left|\cos\theta\right|<0.65\rad$ for the other track,
$P_1 > 0.85$, 
$P_2 > 0.75$,
and $\alpha < 20^\circ$.
At least one track must have an associated EMC cluster
with CM energy less than 0.5\gev. If a cluster
is associated to the second track, its CM energy is required
to be less than 1\gev.

Diphoton candidates are selected by requiring events with
two EMC clusters with energies $E_1, E_2$ satisfying
$2 E_1/\sqrt{s} > 0.85$ and 
$2 E_2/\sqrt{s} > 0.75$.
The CM polar angles of the clusters must satisfy $\left|\cos\theta\right|<0.7\rad$
for one cluster and $\left|\cos\theta\right|<0.65\rad$ for the other,
and the acolinearity angle must be smaller than $10^\circ$.
If there are tracks in the event, the track with the largest
CM momentum must satisfy $P_1 < 0.5$.

Hadronic events ($\epem\to {\rm hadrons}$) are used in the estimation
of the $\Upsilon\to\epem$ background. We select such events by
requiring at least three tracks and a primary vertex location consistent with
the known beamspot. The total energy of tracks and clusters must be
greater than $0.3\sqrt{s}$, and the ratio of the second to the zeroth
Fox-Wolfram moments~\cite{ref:R2} is required to be smaller than 0.95.
The distance between the primary production vertex of the tracks in the
event and the time-averaged beamspot position must be less than 0.5~cm
in $r$
and less than 6~cm in $z$.

\begin{figure*}[htp!]
\begin{tabular} {lr}

\includegraphics[width=0.5 \linewidth]{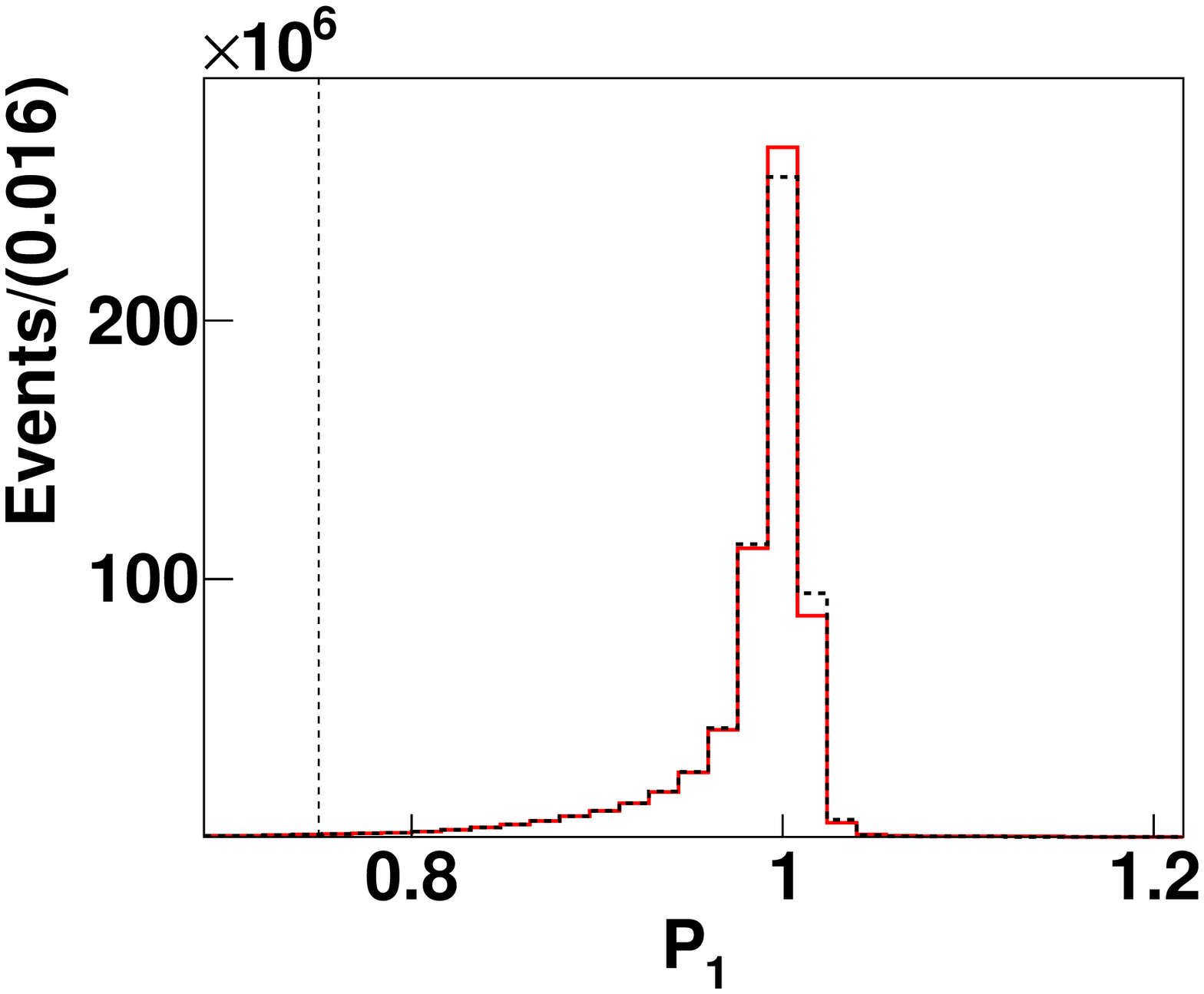} &
\includegraphics[width=0.5 \linewidth]{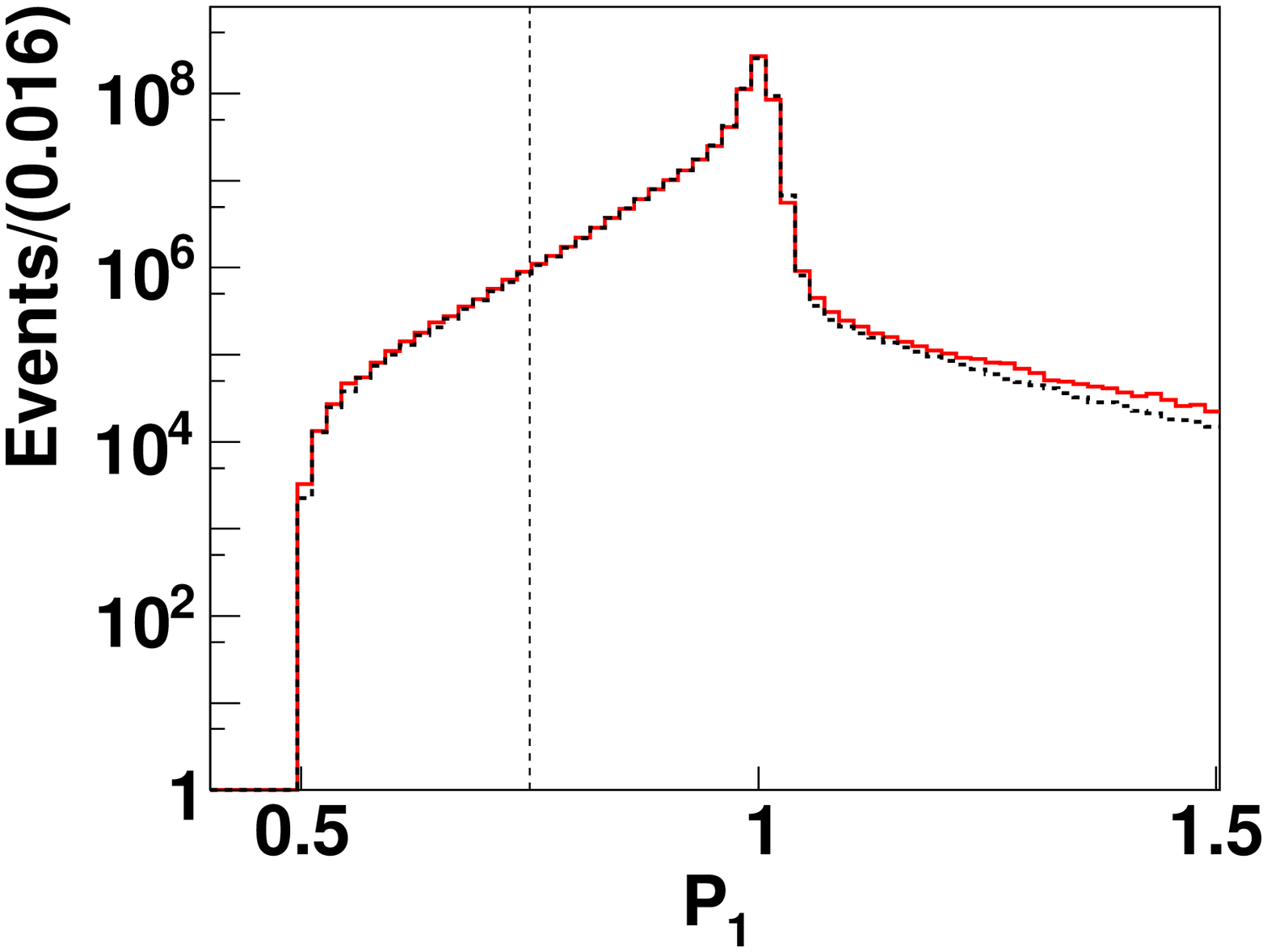}  \\
\includegraphics[width=0.5 \linewidth]{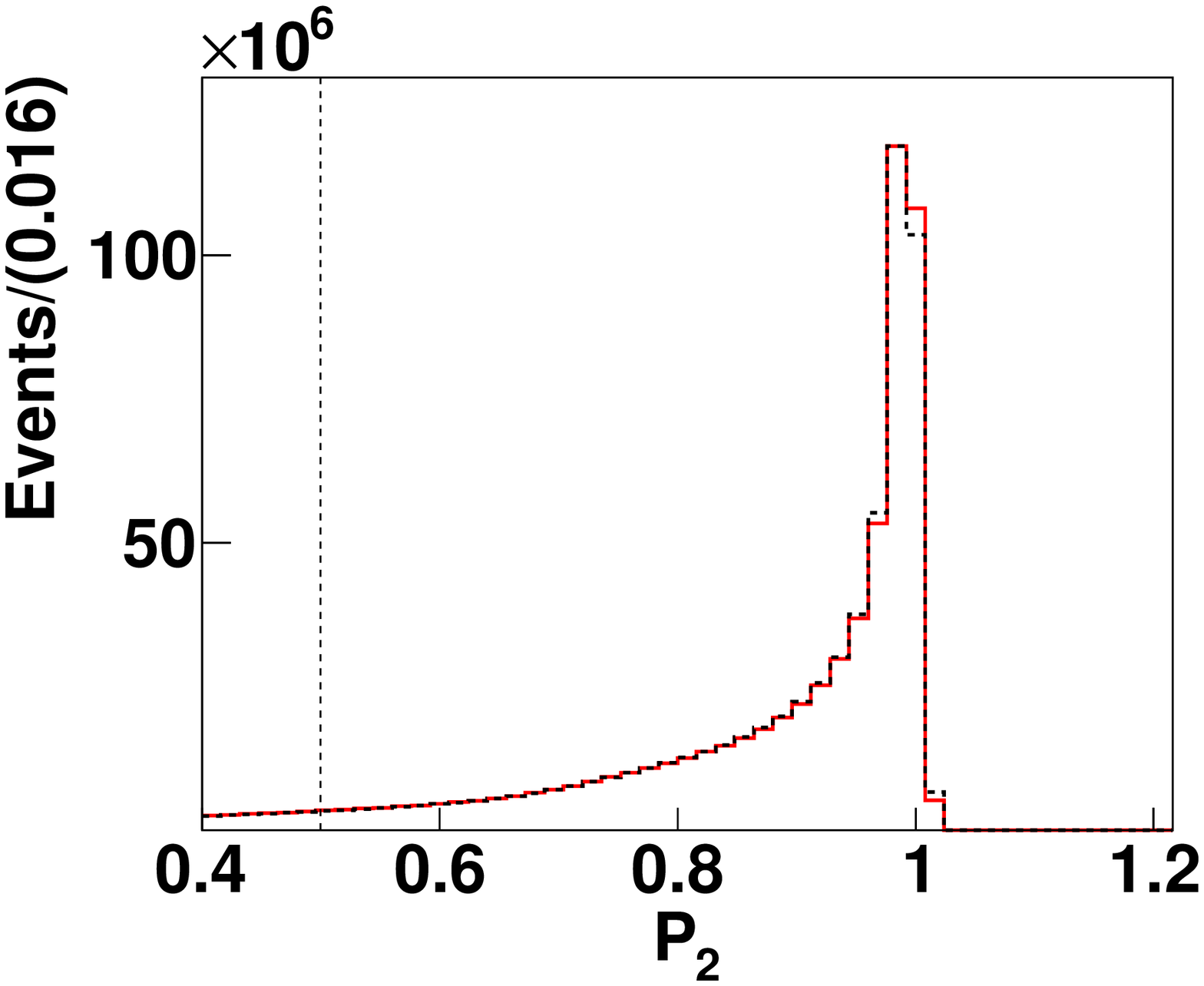} &
\includegraphics[width=0.5 \linewidth]{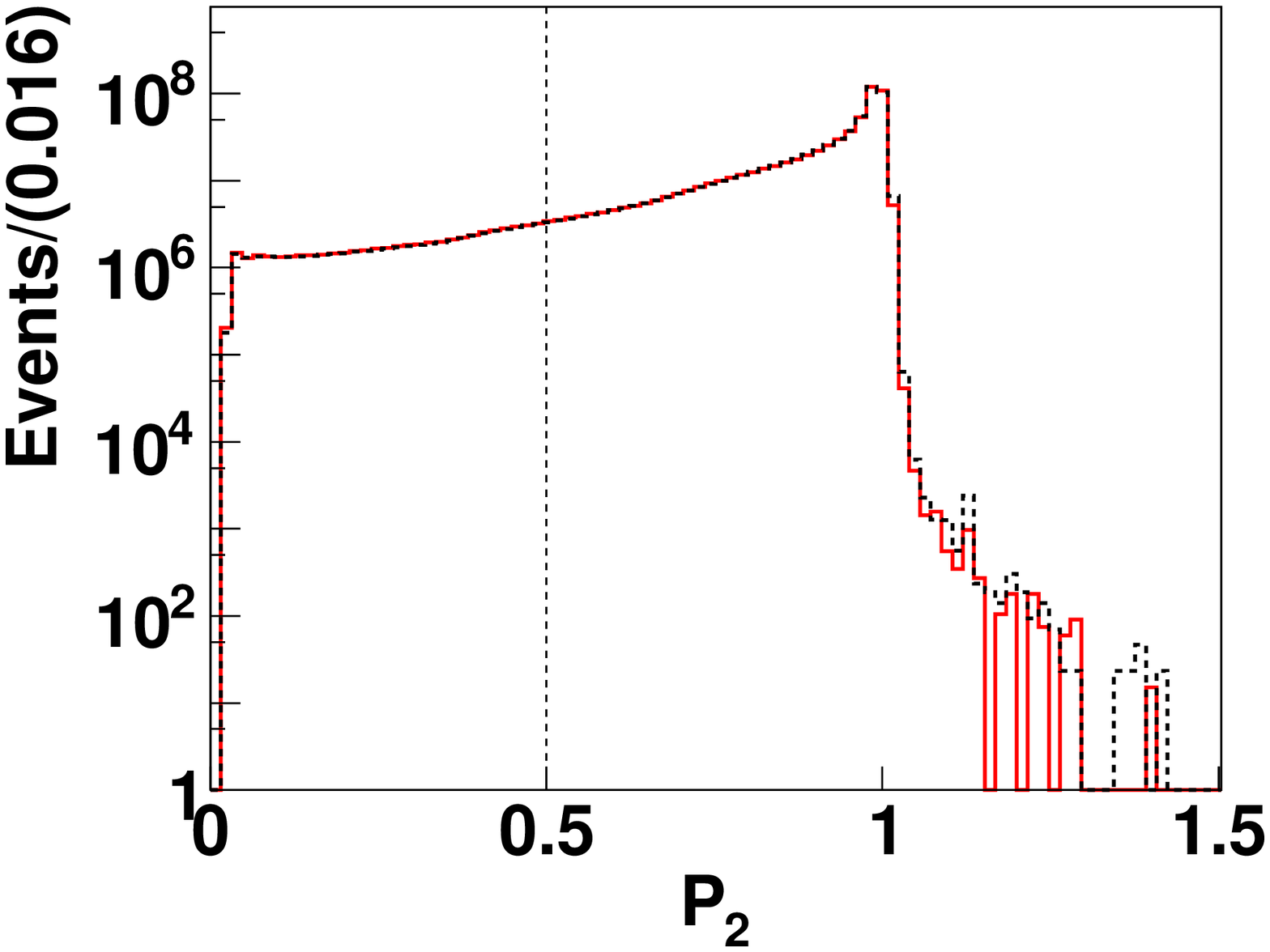}  \\
\includegraphics[width=0.5 \linewidth]{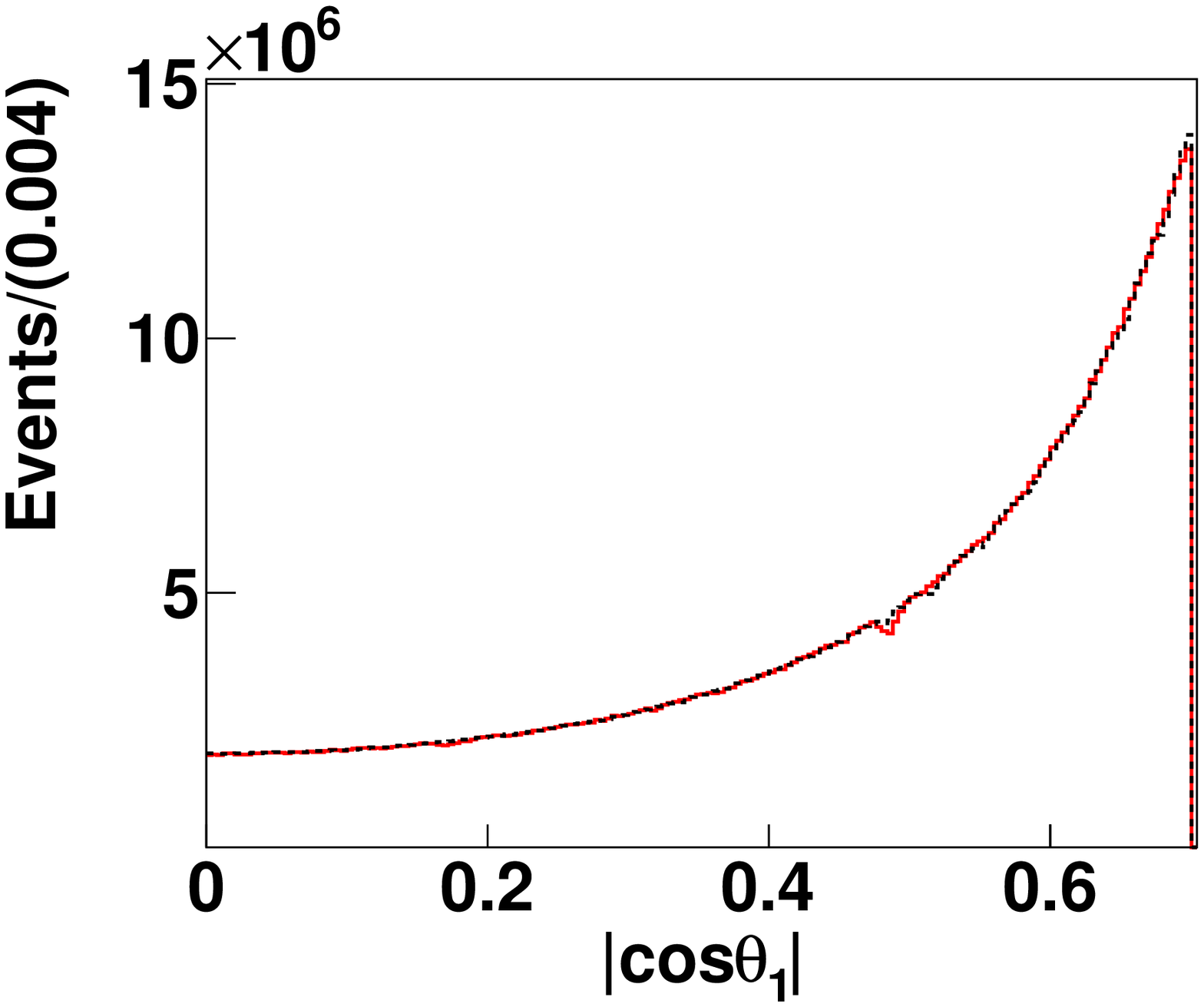}   &
\includegraphics[width=0.5 \linewidth]{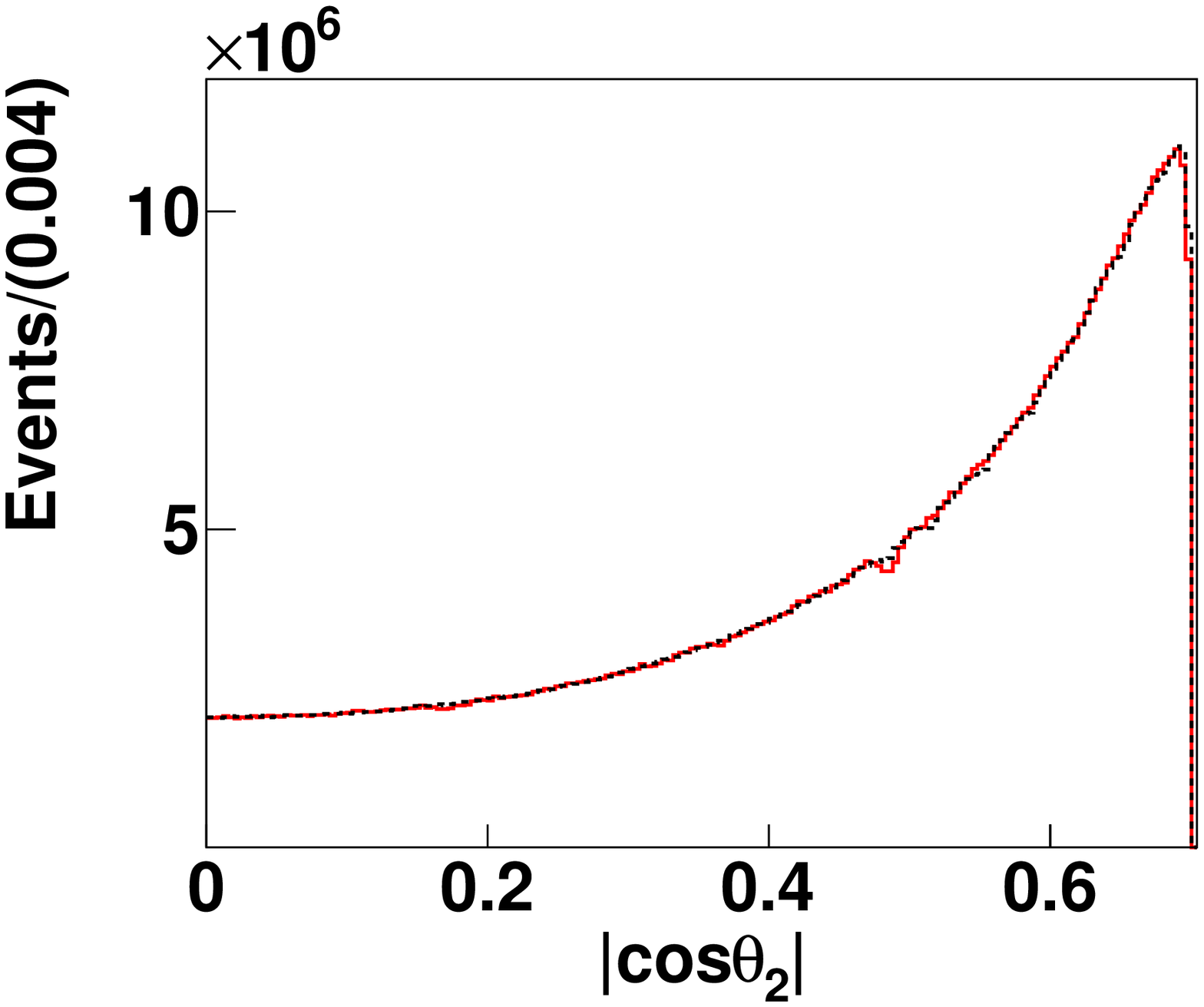}   \\
					
\end{tabular}
\caption{\label{figfirst} 
Distributions of the scaled CM momentum 
$P_i = 2 p_i / \sqrt{s}$ 
and cosine of the CM polar angle
$\theta_i$ for the 
higher-momentum ($i=1$) and lower-momentum ($i=2$) track in
candidate $\eeMode$ events in a fraction of the
data (Run 4; solid red histograms) and for simulated 
$\eeMode$ events (dashed black histograms).
The simulation histograms are normalized to the area of the data
histograms.  The upper two rows of figures show the $P_i$ distributions
with linear (left) and log (right) vertical scale. In each
scaled-momentum
plot, the vertical line shows the minimum value for events that
are retained. When plotting each variable, the selection criteria on
all other variables are applied. The $\left|\cos\theta_i\right|$ ($i=1,2$) 
plots are made
with $\left|\cos\theta_i\right| < 0.7$.}
\end{figure*}

\begin{figure*}
\begin{tabular} {lr}
\includegraphics[width=0.5 \linewidth]{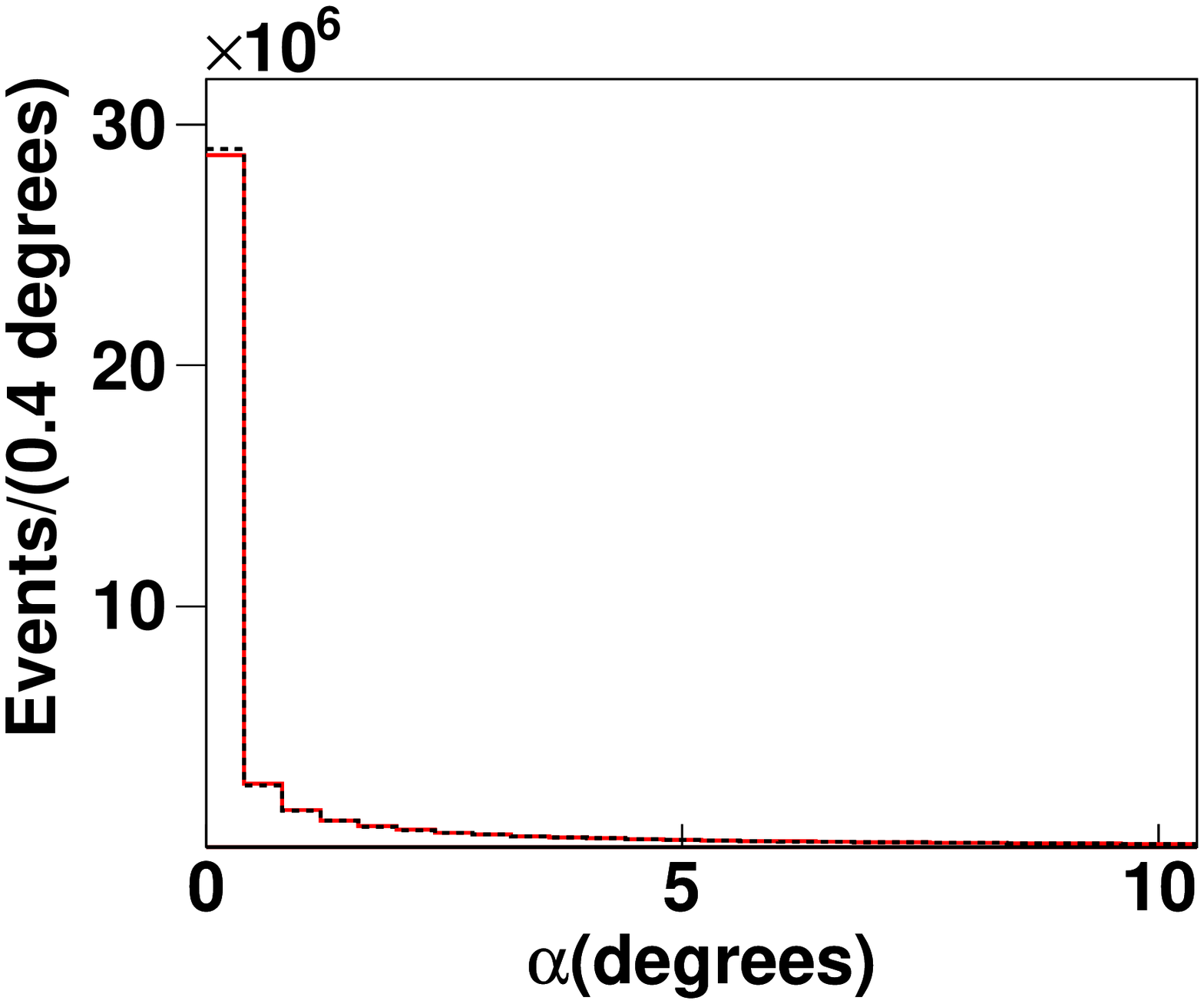} &
\includegraphics[width=0.5 \linewidth]{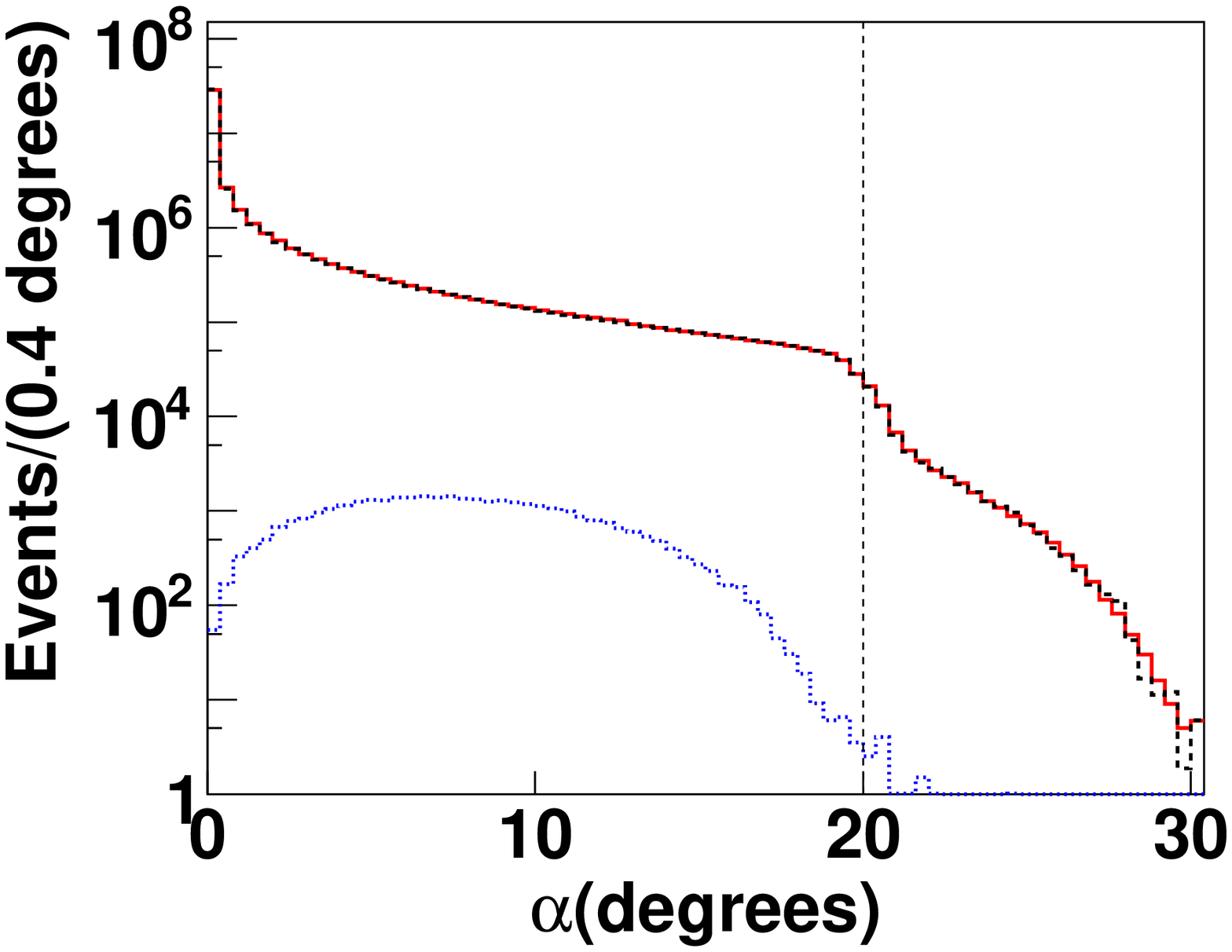}\\
\includegraphics[width=0.5 \linewidth]{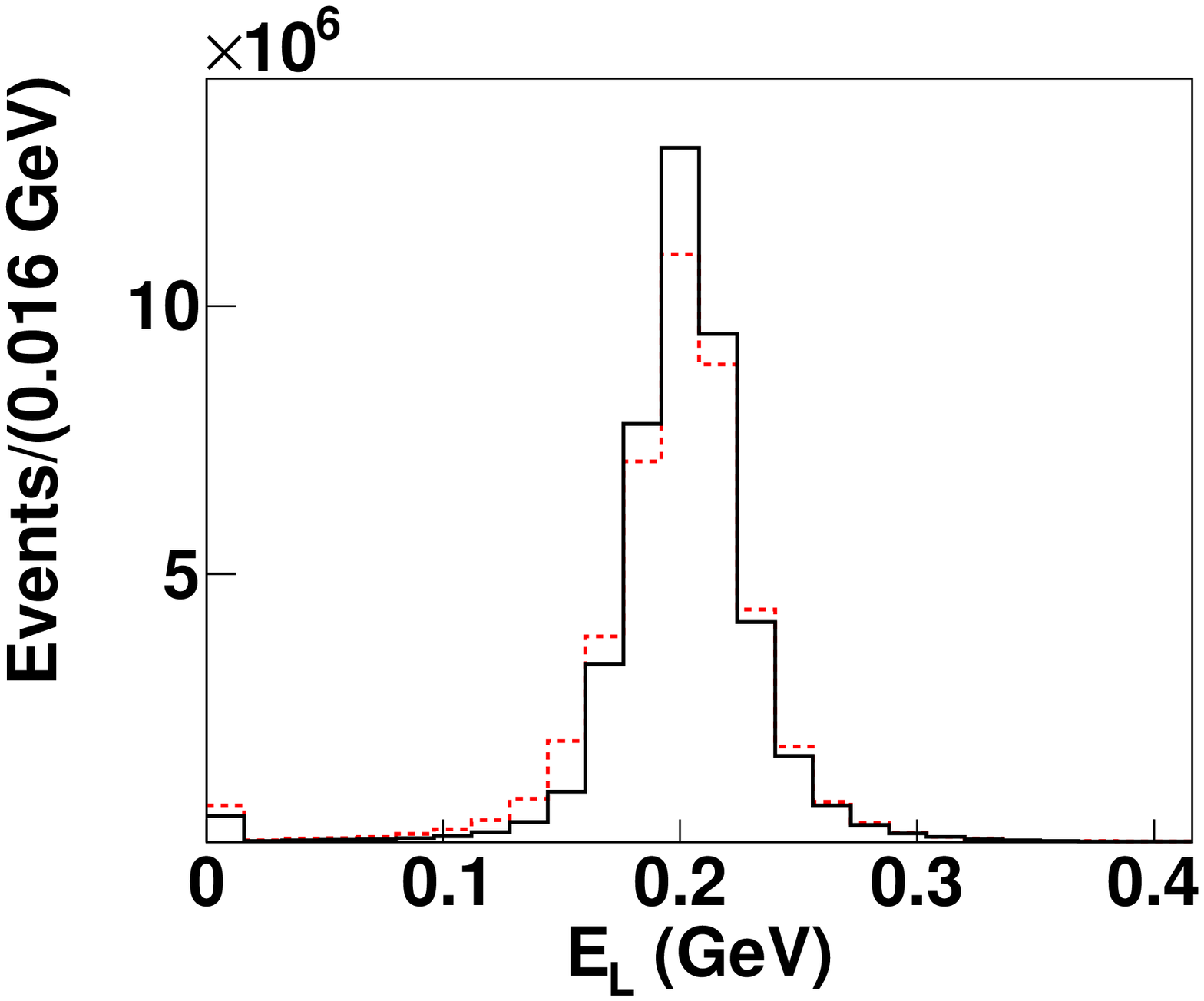}   &
\includegraphics[width=0.5 \linewidth]{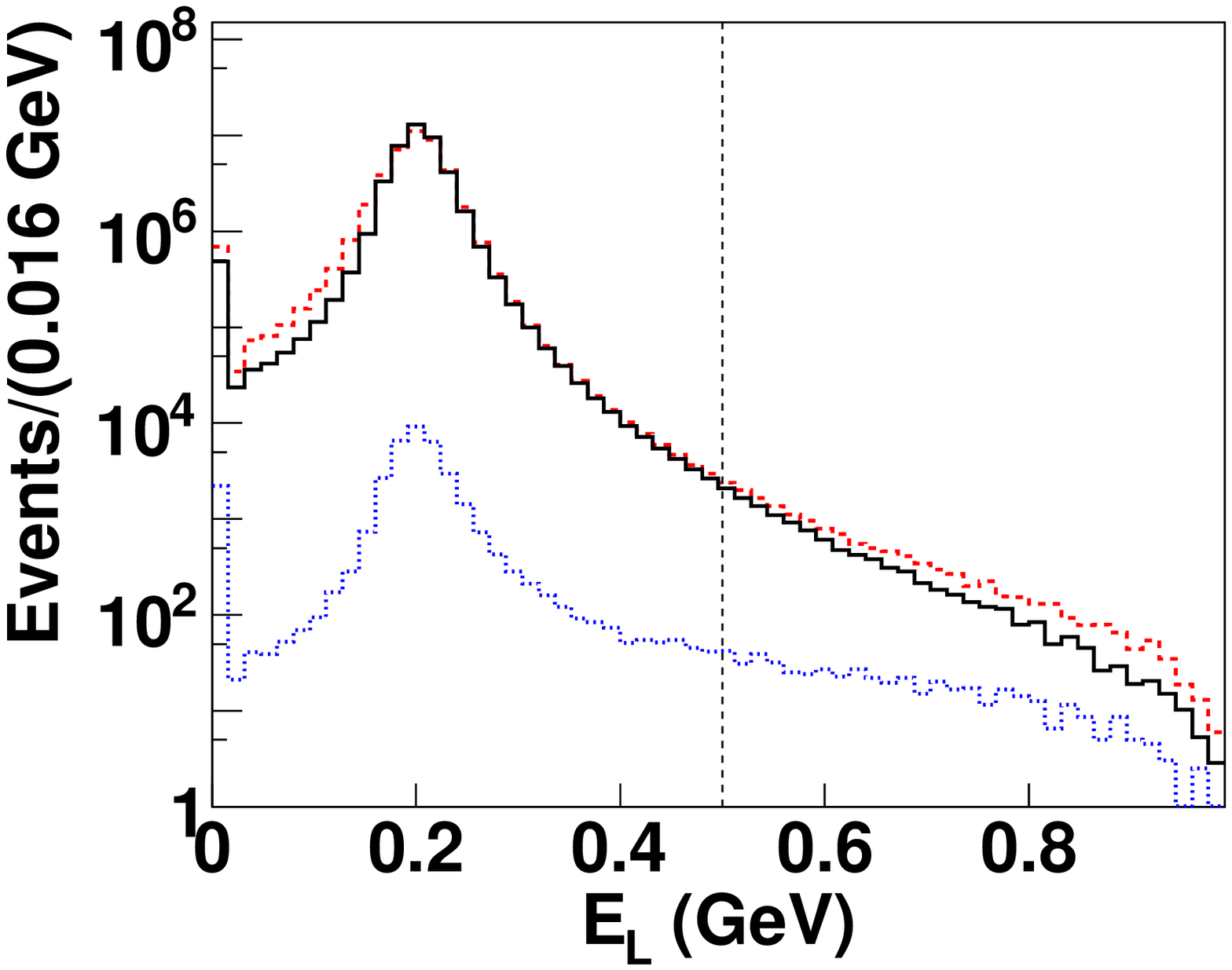}   \\ 
\includegraphics[width=0.5 \linewidth]{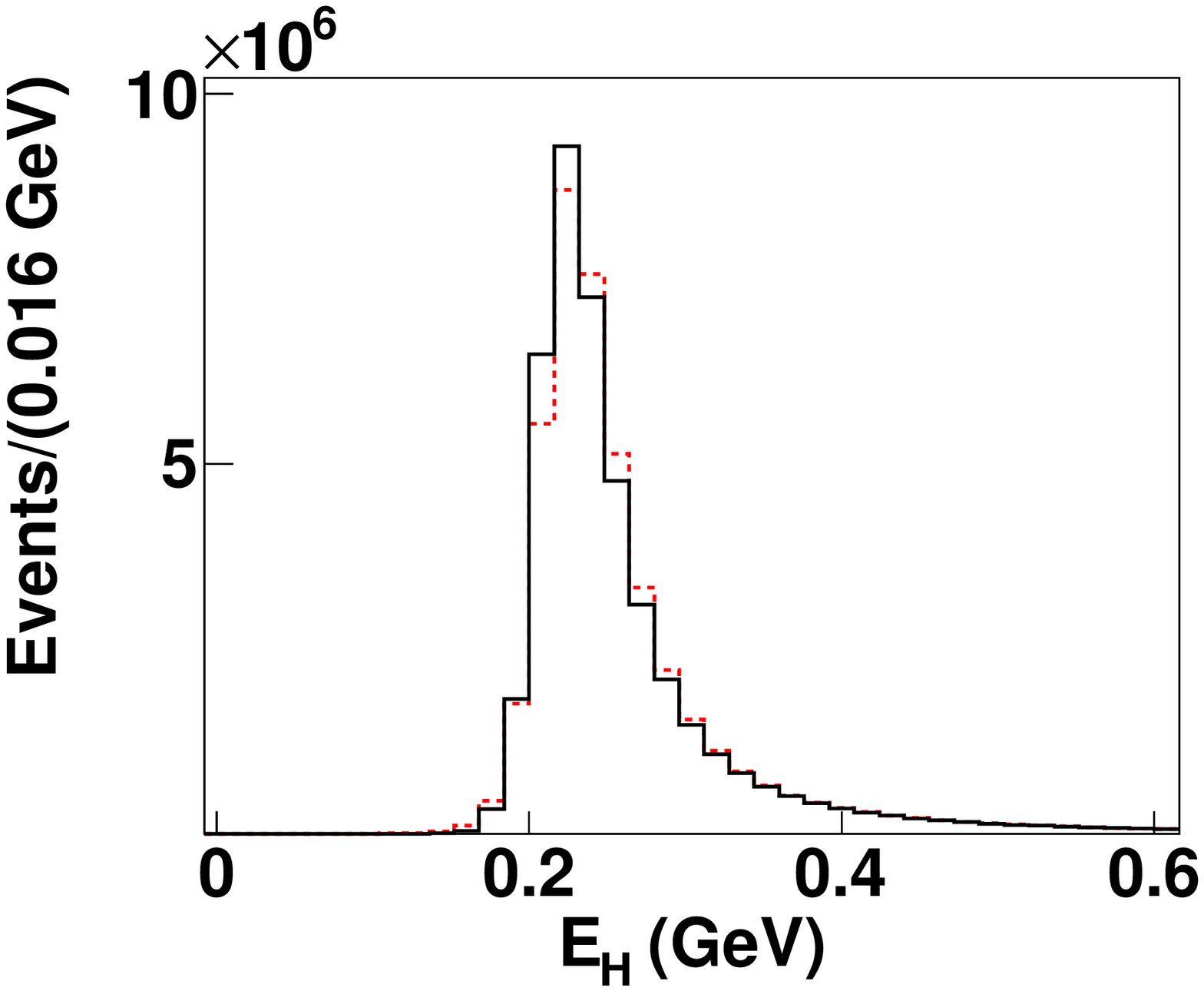}   &
\includegraphics[width=0.5 \linewidth]{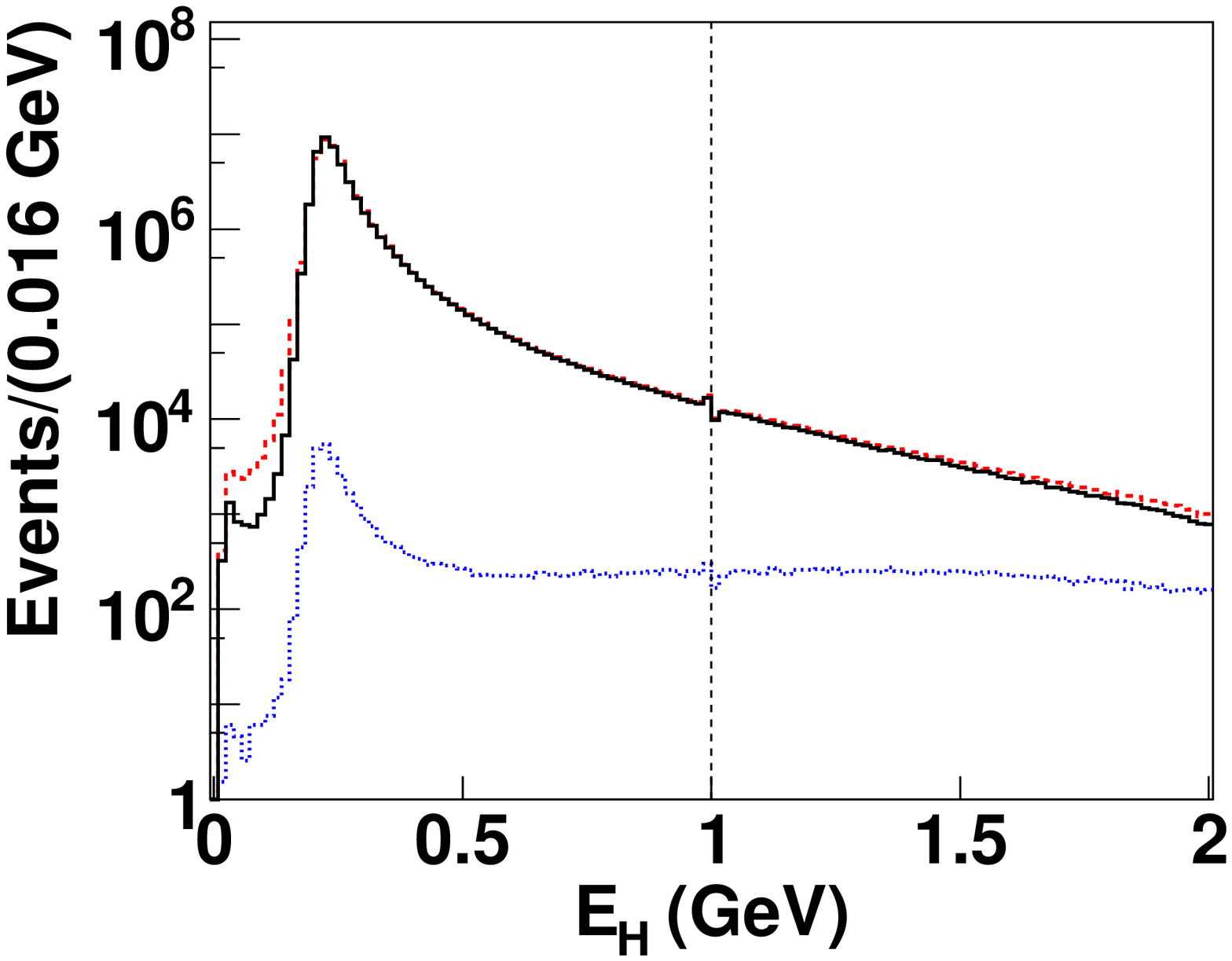} \\
					
\end{tabular}
\caption{\label{figlast} 
Distributions of the CM acolinearity angle $\alpha$
and of the laboratory-frame energies of the higher-energy
($E_H$) and lower-energy ($E_L$) EMC clusters matched to the tracks in
candidate \mumuMode\ events in a fraction of the
data (Run 4; solid red histograms) and for simulated 
$\mumuMode$ and \tautauMode\ events (dashed black histograms).
In each log-scale plot,
the dotted blue histogram shows the small contribution of
\tautauMode\ events to the simulation histogram,
and the vertical line 
shows the maximum
value for events that are retained.
The simulation histograms are normalized to the area of the data
histograms. The distributions are shown with linear (left) and log
(right) vertical scale. When plotting each variable, the selection criteria on
all other variables are applied.
The small structure at $1\gev$ in the $E_H$ distribution results from a
high-energy calibration correction that is applied to clusters with
$E>1\gev$.}
\end{figure*}

Figs.~\ref{figfirst} through~\ref{figlast} show examples of the Bhabha
and dimuon selection-variable distributions for data and simulation.
Although in some cases there are visible differences
between the distributions in data and in simulation, the loose
selection criteria ensure that these differences have negligible
impact on the knowledge of the signal efficiency.


\begin{figure*}
\begin{tabular} {lr}

\includegraphics[width=0.5 \linewidth]{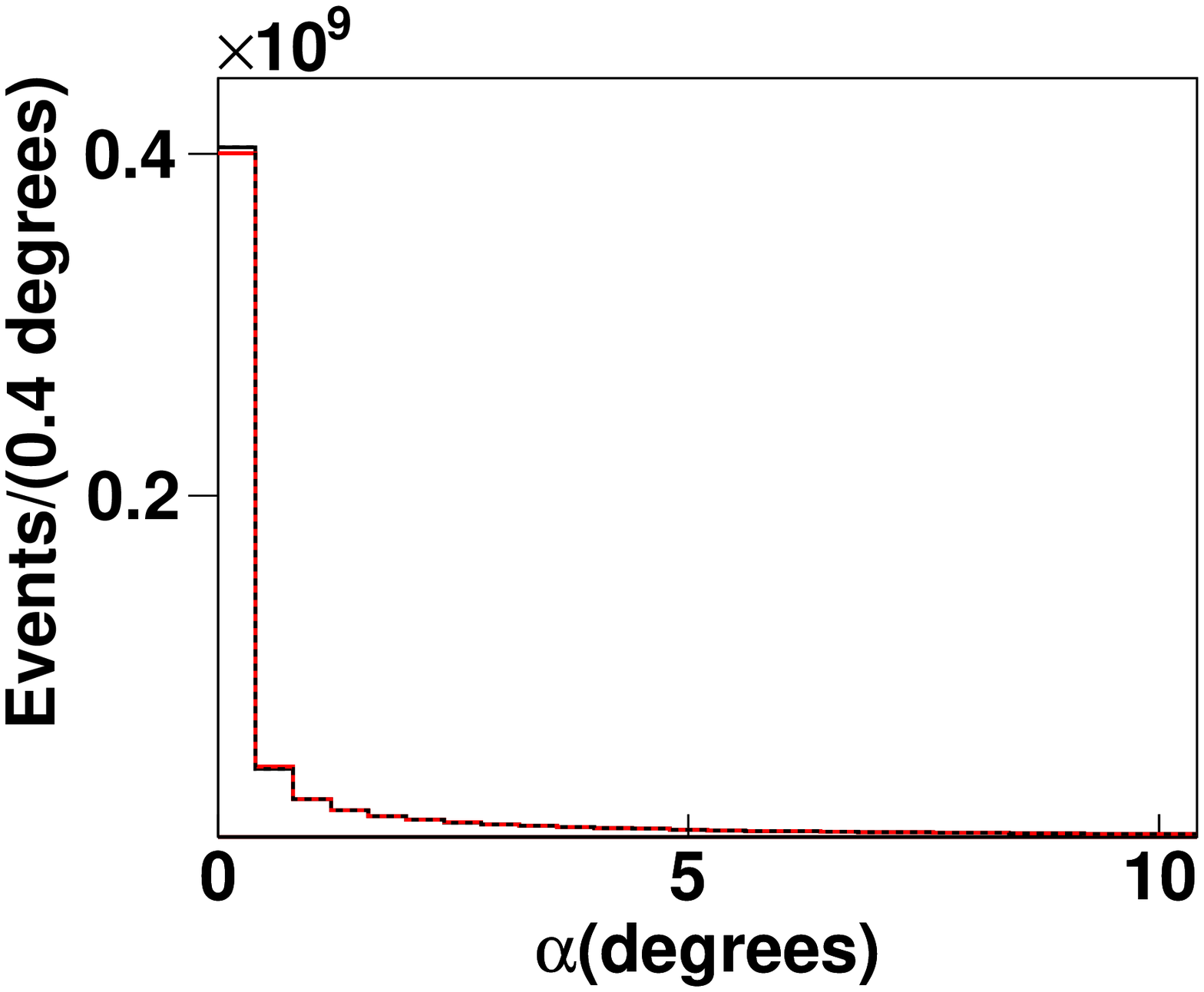}&
\includegraphics[width=0.5 \linewidth]{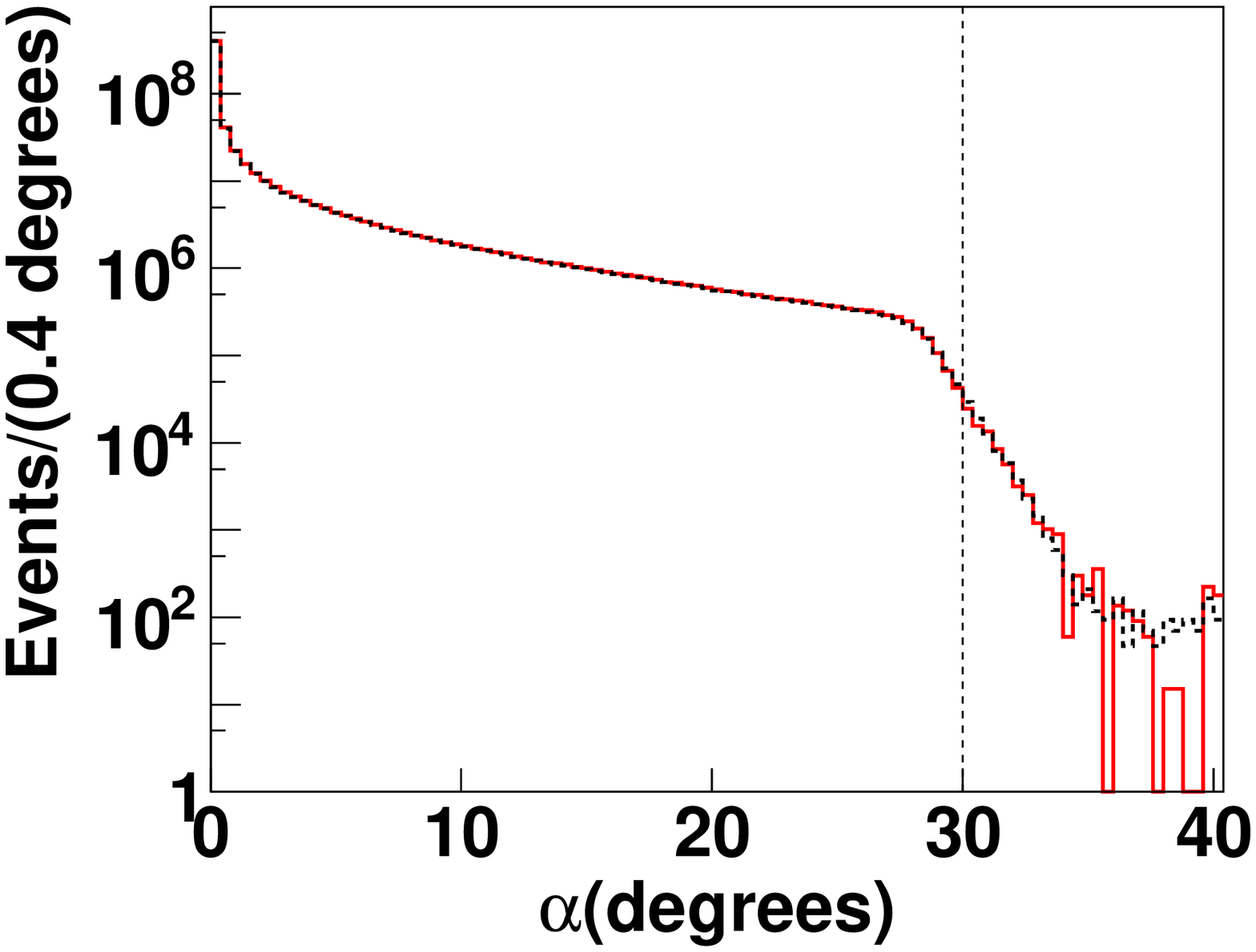} \\
\includegraphics[width=0.5 \linewidth]{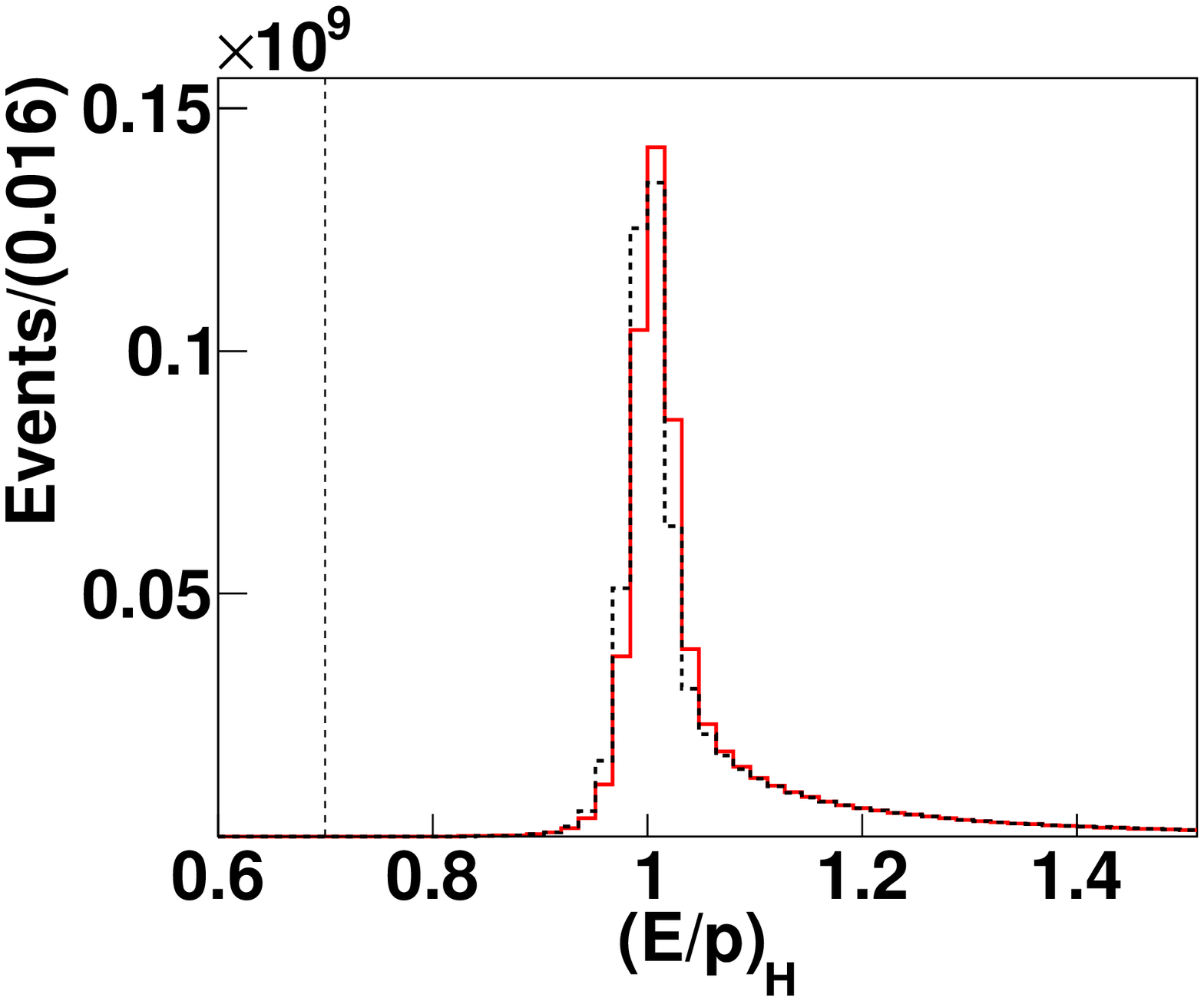} &
\includegraphics[width=0.5 \linewidth]{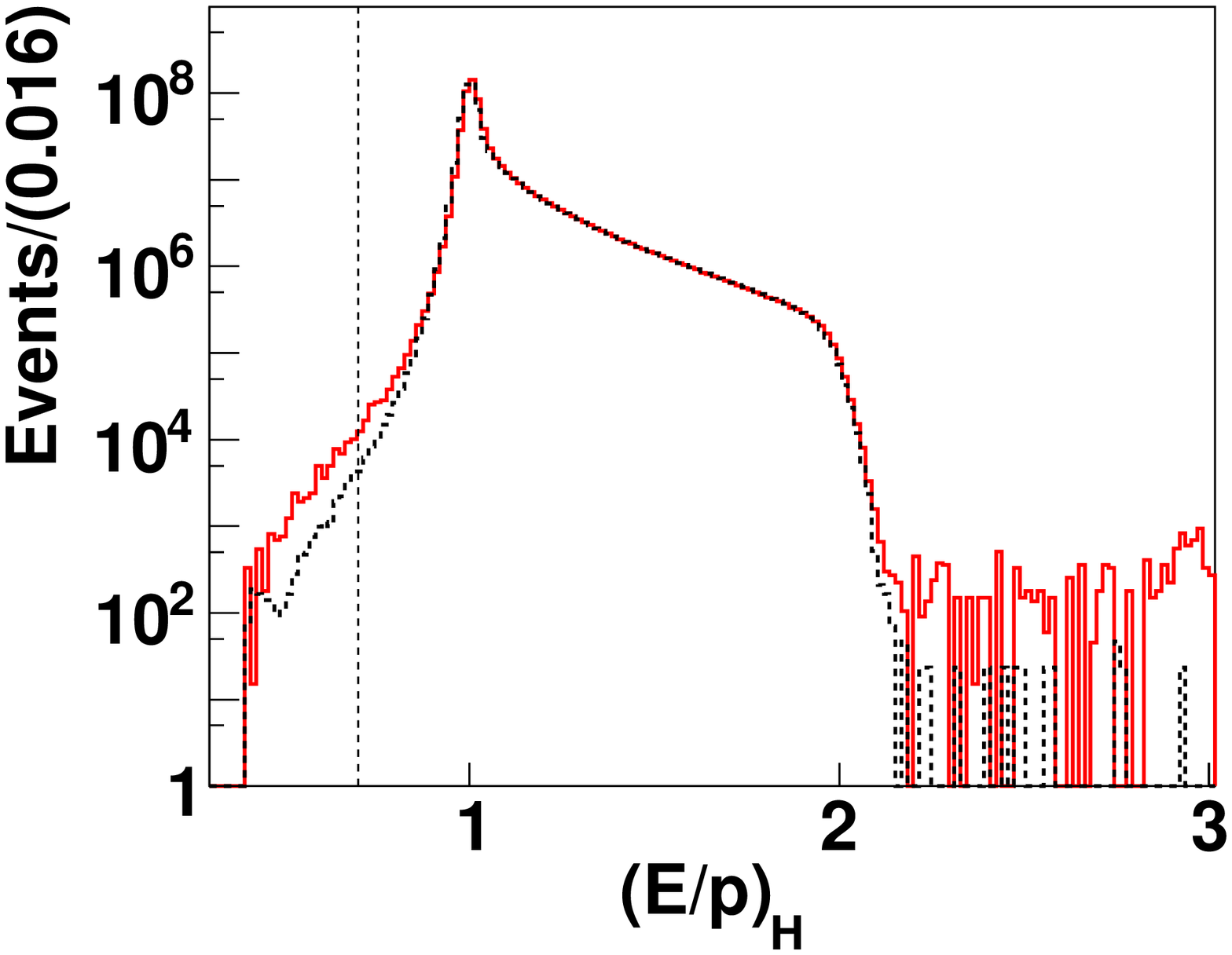}   \\
\includegraphics[width=0.5 \linewidth]{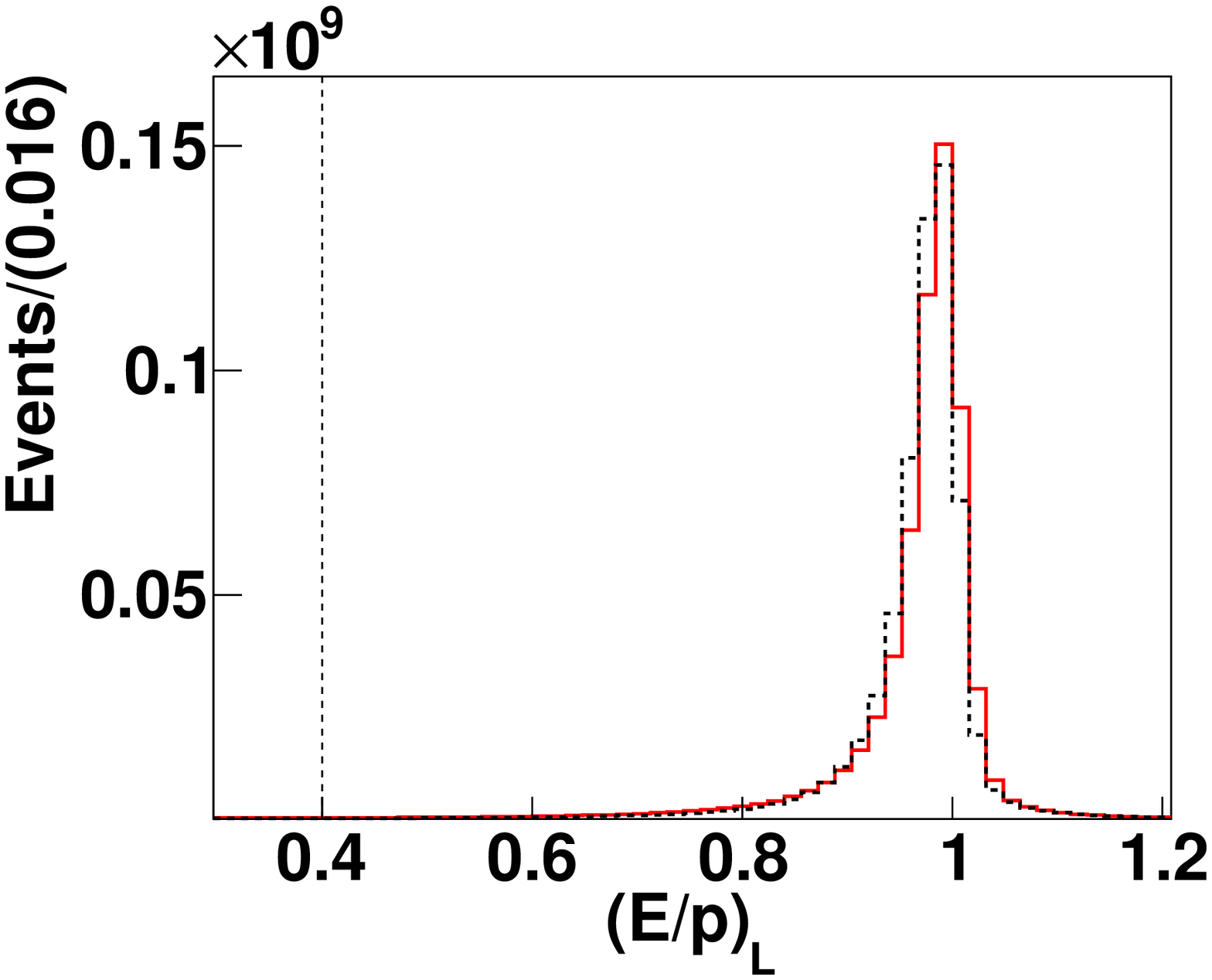} &
\includegraphics[width=0.5 \linewidth]{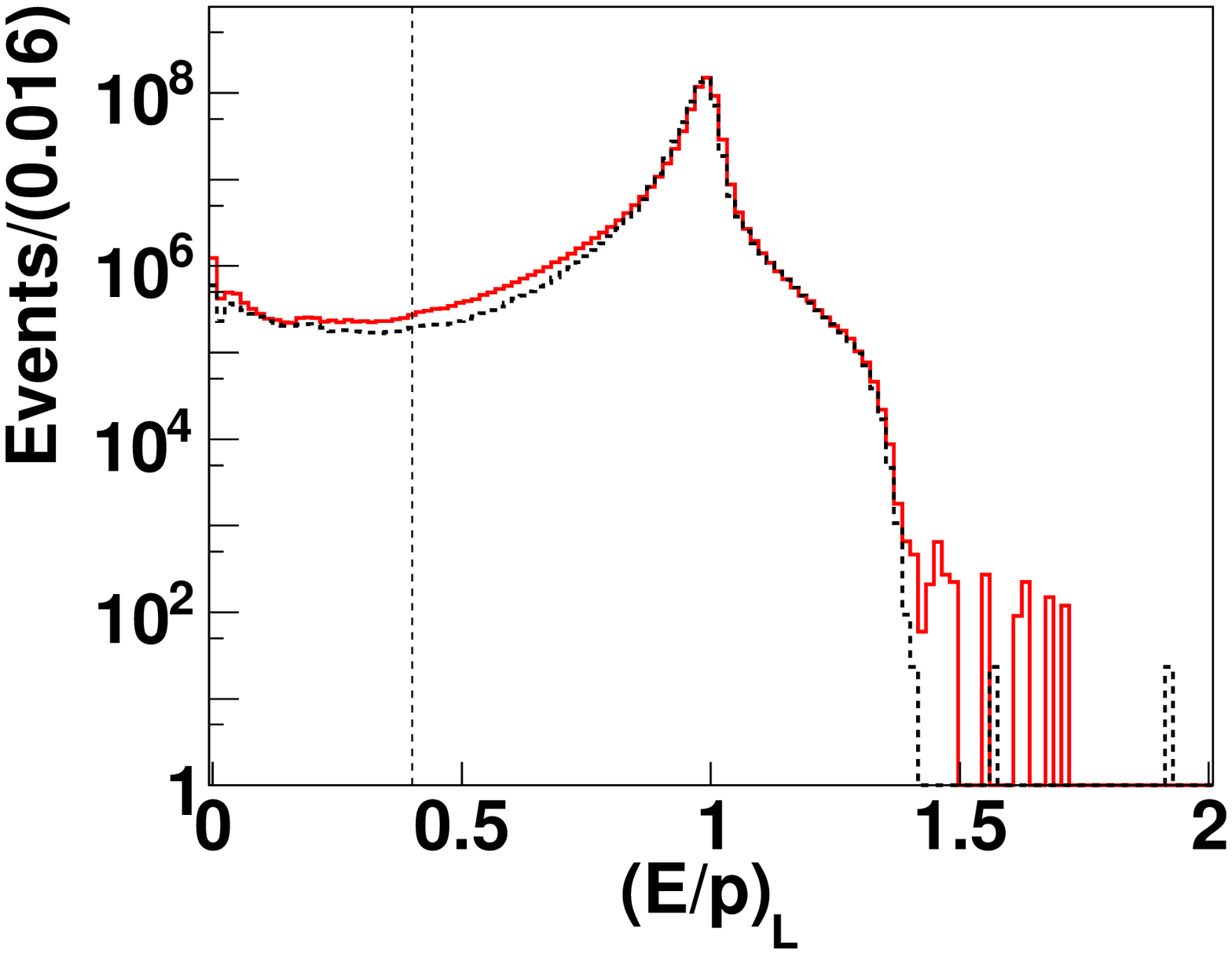} \\
					
\end{tabular}
\caption{Distributions of the CM acolinearity angle $\alpha$,
and the higher (lower) laboratory-frame energy-to-momentum ratio 
$E/p_H$ ($E/p_L$) for \eeMode\ candidates in a fraction of the
data (Run 4; solid red histograms) and for simulated 
$\eeMode$ events (dashed black histograms).
The simulation histograms are normalized to the area of the data
histograms. The distributions are shown with linear (left) and log
(right) vertical scale. In each $E/p$ plot (log-scale $\alpha$ plot), the vertical line 
shows the minimum (maximum)
value for events that are retained.
When plotting each variable, the selection criteria on
all other variables are applied.}
\end{figure*}

\begin{figure*}
\begin{tabular} {lr}

\includegraphics[width=0.5\linewidth]{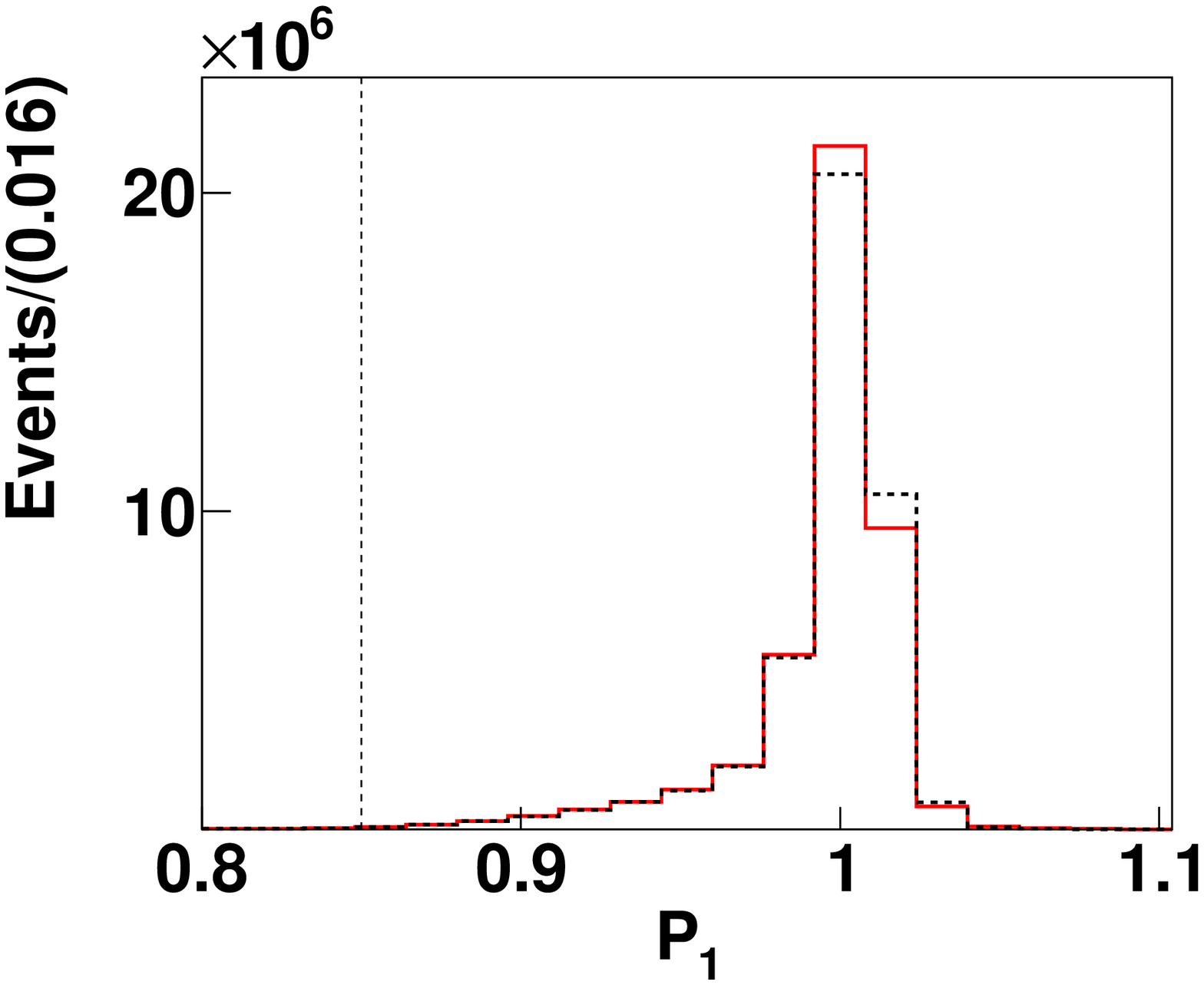}&
\includegraphics[width=0.5 \linewidth]{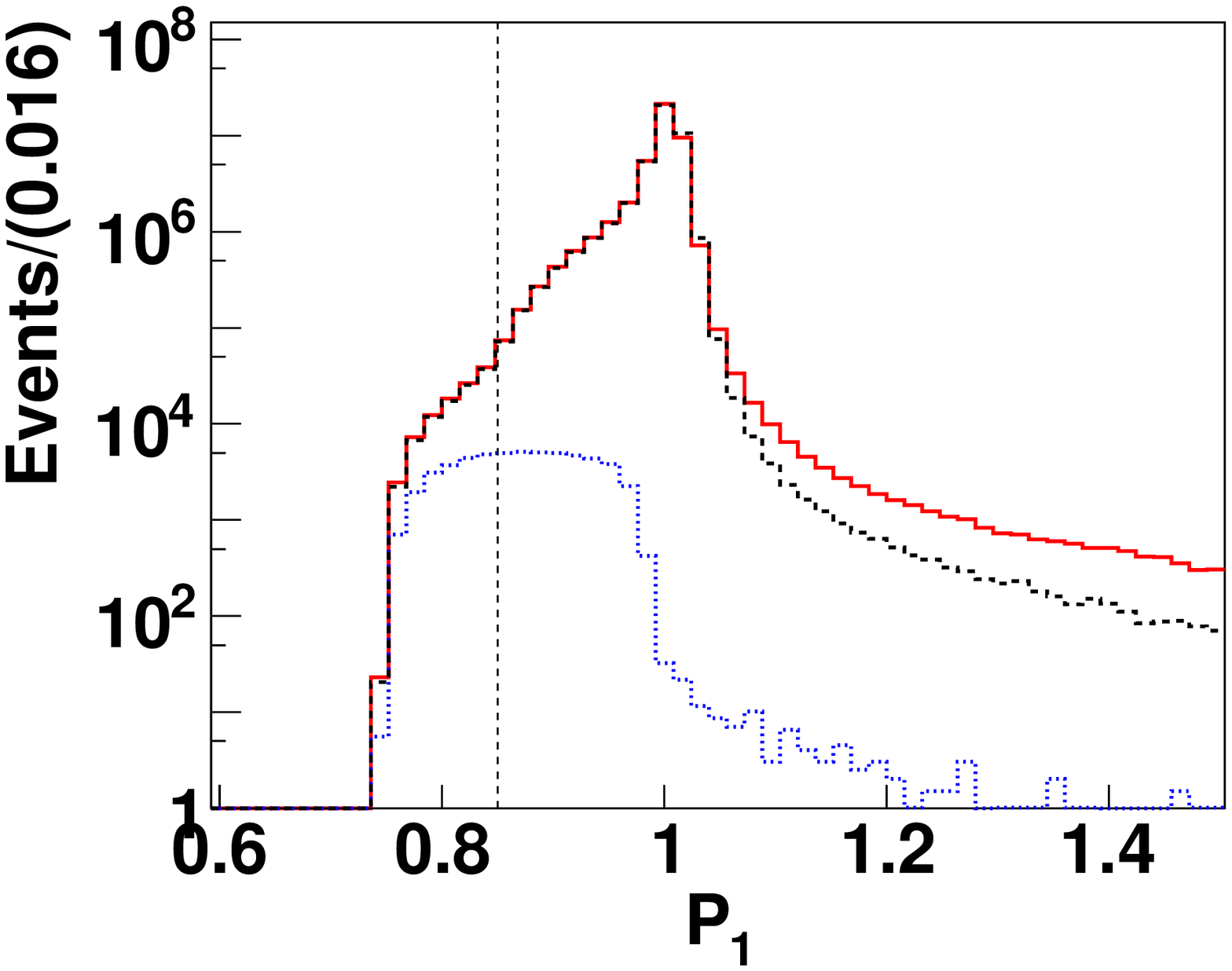} \\
\includegraphics[width=0.5\linewidth]{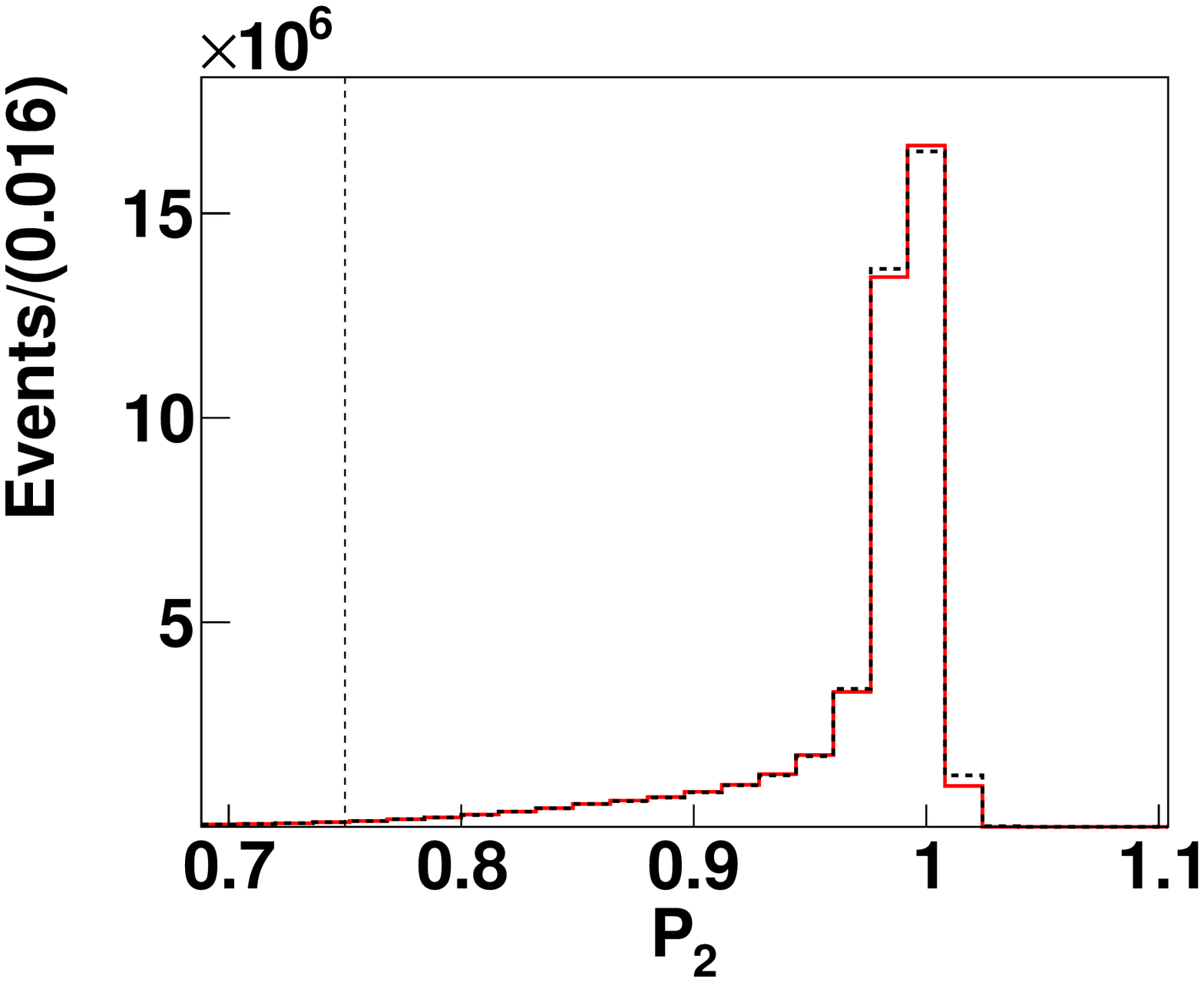}&
\includegraphics[width=0.5 \linewidth]{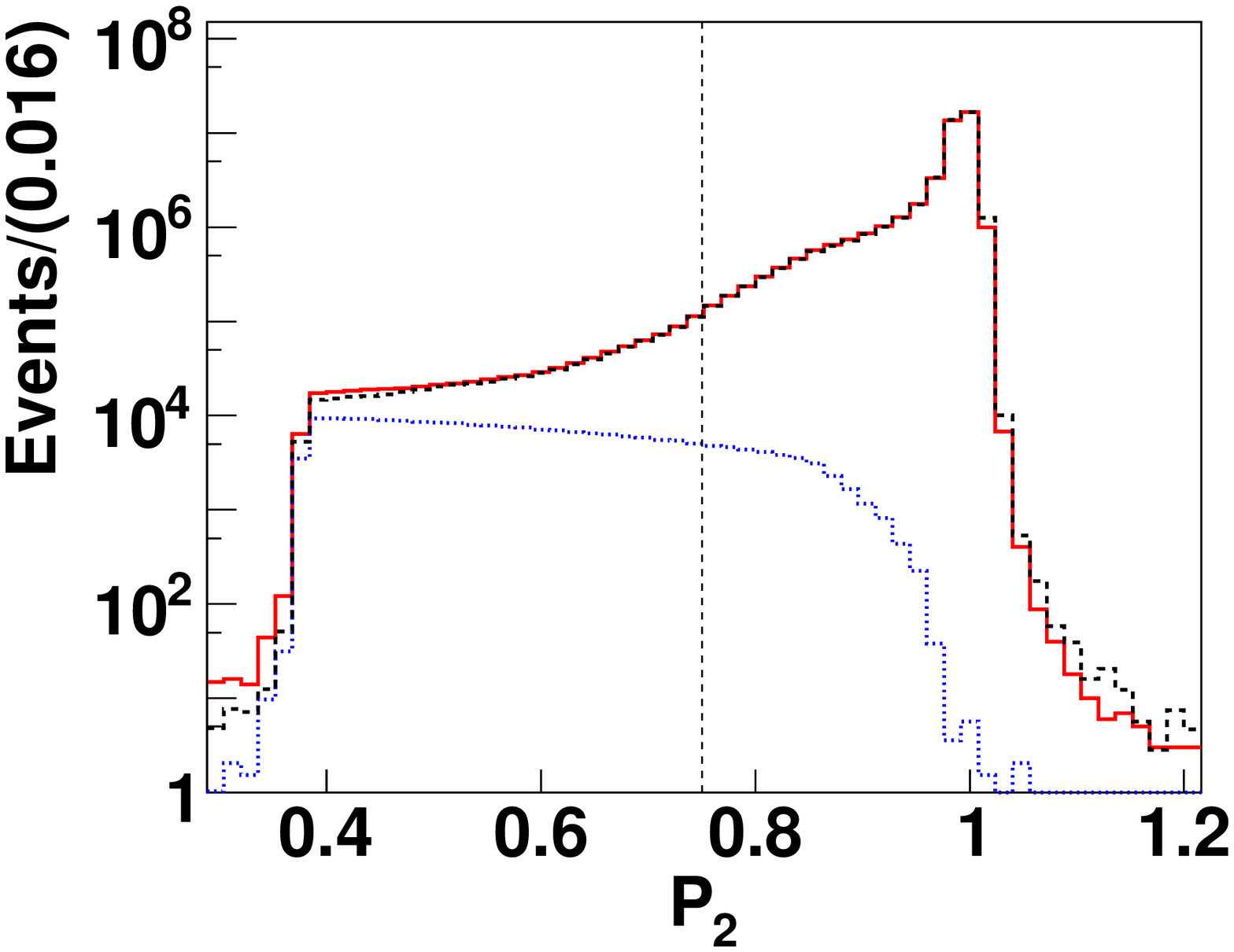} \\
\includegraphics[width=0.5 \linewidth]{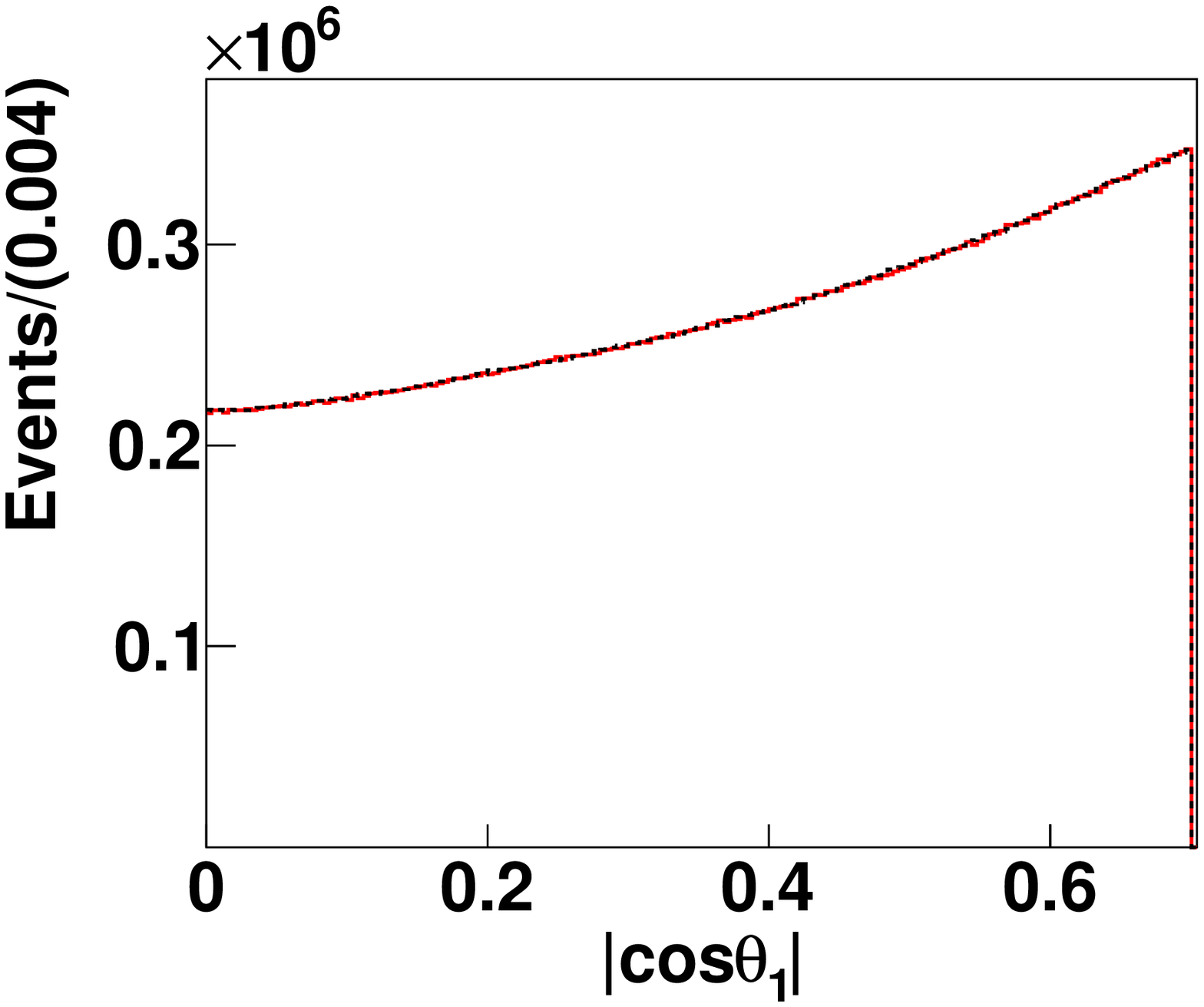}   &
\includegraphics[width=0.5 \linewidth]{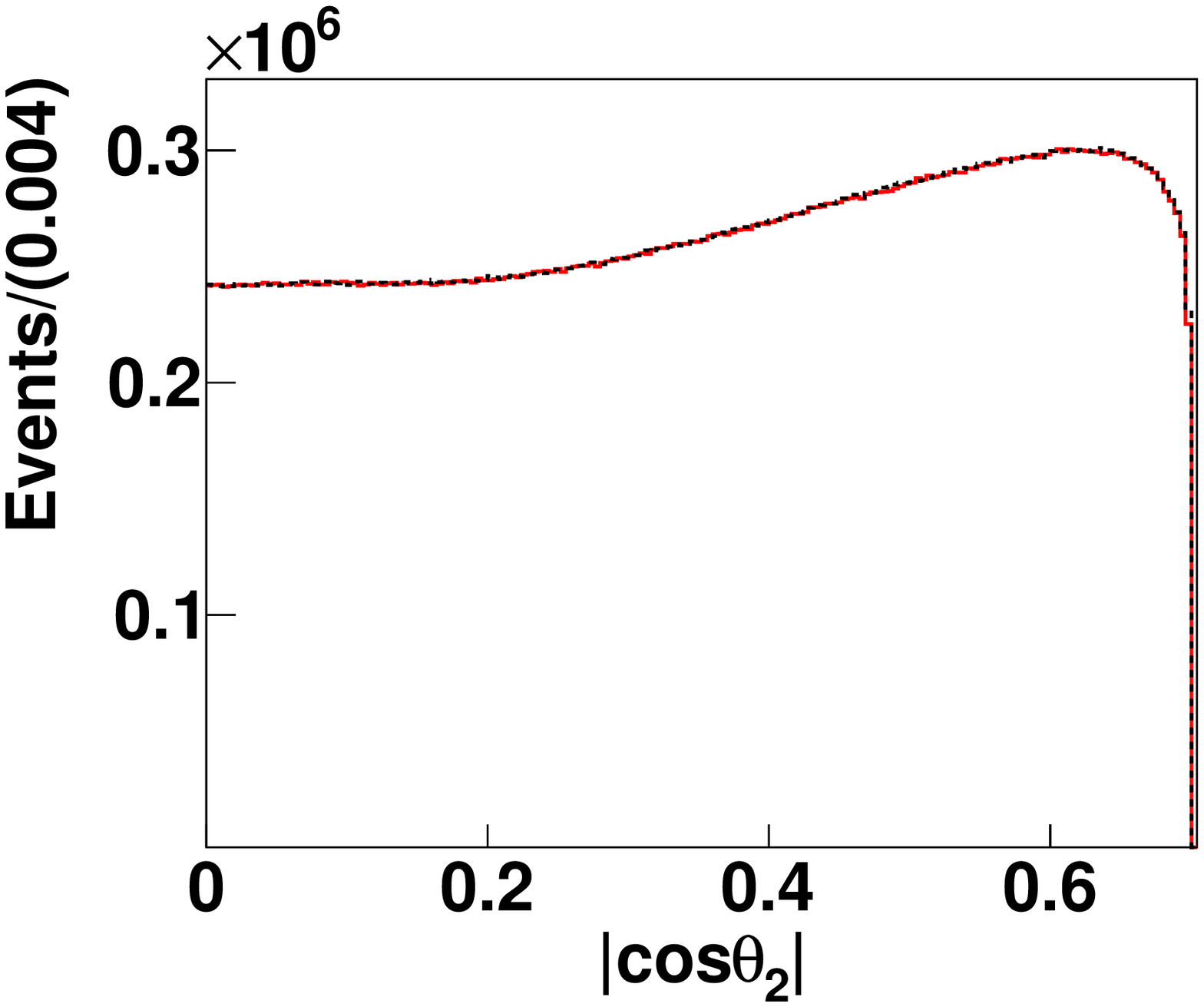}   \\ 
\end{tabular}
\caption{Distributions of the scaled CM momentum 
$P_i = 2 p_i / \sqrt{s}$ 
and cosine of the CM polar angle
$\theta_i$ for the 
higher-momentum ($i=1$) and lower-momentum ($i=2$) track in
candidate $\mumuMode$ events in a fraction of the
data (Run 4; solid red histograms) and for simulated 
$\mumuMode$ and \tautauMode\ events (dashed black histograms).
In the log-scale plots, 
the dotted blue histograms show the small contribution of
\tautauMode\ events to the simulation histograms.
The simulation histograms are normalized to the area of the data
histograms.  The upper two rows of figures show the $P_i$ distributions
with linear (left) and log (right) vertical scale. 
In each
scaled-momentum
plot, the vertical line shows the minimum value for events that
are retained. When plotting each variable, the selection criteria on
all other variables are applied. The $\left|\cos\theta_i\right|$ ($i=1,2$) 
plots are made
with $\left|\cos\theta_i\right| < 0.7$.}
\end{figure*}

\subsection{Background Estimation}
\label{sec:bgd}

\subsubsection{Background Sources Common to All Runs}


The efficiency for $\tautauMode \to \mu^+
\mu^- \nu_\mu \bar\nu_\mu \nu_\tau \bar\nu_\tau$ events to pass the
dimuon selection is determined using MC. We find the fraction of such events
in the selected $\mumuMode$ candidate sample to be $(0.0816 \pm 0.0033)\%$.
The fraction of Bhabha events in the dimuon sample is 
determined in the same way, and is found to be $(0.02 \pm 0.01)\%$.
In both cases, the uncertainties are due to MC statistics, and are much larger
than those expected due to uncertainties on the efficiency or
the cross sections of the various modes.

To estimate the background due to cosmic rays or beam-gas
interactions, we select dimuon candidates where the point of closest
approach of the tracks to the beamline is between 10~cm and 30~cm of
the interaction point in $z$,
and that satisfy all other requirements.  From this sample, the level
of contamination of cosmic events in the dimuon sample is determined to
be $(1.8 \pm 0.7) \times 10^{-5}$, which we take to be negligible.

The background level in the Bhabha sample is much smaller than the
values listed above for the dimuon sample, since the visible cross
section for \eeMode\ is an order of magnitude larger than for \mumuMode.
Therefore, the background in the Bhabha channel is neglected.

\subsubsection{$\Upsilon$ Background in Run~7}

The on-resonance Run-7 sample
contains non-negligible contributions 
from the decays
$\Upsilon(2S)\to \epem$, $\Upsilon(3S)\to\epem$ and,
to a smaller extent, from
cascade decays such as 
$\Upsilon(2S) \to \pi^+\pi^- \Upsilon(1S) \to \pi^+\pi^- \epem$ or 
$\Upsilon(2S) \to \gamma \chi_{bJ}(1P) \to \gamma \gamma \Upsilon(1S)
\to \gamma \gamma \epem$.
This type of background, which we label as 
$\Upsilon \to \epem X$, is negligible in the $\Upsilon(4S)$ samples 
of Runs~1--6.
We determine the number of Run-7 $\Upsilon\to\epem X$ events from 
\beq
N_{\Upsilon\to\epem} = N_{\Upsilon} \BR_{\rm vis}(\Upsilon\to\epem X),
\label{eq:nUps-to-ee}
\eeq
where  $N_{\Upsilon}$ is the number of $\epem\to\Upsilon$ events produced,
and the visible branching fraction 
\beqa
&& \BR_{\rm vis}(\Upsilon\to\epem X) = \nonumber\\
&&~~  \sum_i \BR_i(\Upsilon\to\epem X)\epsilon_i(\Upsilon\to\epem X)
\eeqa
accounts for the branching fraction $\BR_i(\Upsilon\to\epem X)$ and
reconstruction efficiency $\epsilon_i(\Upsilon\to\epem X)$ of each
process (indicated by the index $i$) that contributes to this
background.
We obtain $\BR_{\rm vis}(\Upsilon\to\epem X)$ from simulated events,
generated with branching fractions $\BR_i(\Upsilon\to\epem X)$
based on the measurements compiled in the 
Review of Particle Physics~\cite{ref:pdg}.  
Since $\BR(\Upsilon(3S)\to\epem)$ has not been measured, we take its
value to be identical to $\BR(\Upsilon(3S)\to\mu^+\mu^-)$, relying on
lepton universality in electromagnetic interactions.
Since the spread in $\sqrt{s}$ (about $5\mev$) is much larger than
the widths of the $\Upsilon(2S)$ and $\Upsilon(3S)$ resonances, we 
ignore interference between $\Upsilon\to\epem$ decays and Bhabha scattering
when estimating the $\Upsilon\to\epem$ background. Interference
is further suppressed by the different polar-angle distributions of the 
two processes.


To determine the number of $\Upsilon$ mesons produced in the 
$\Upsilon(2S)$ or $\Upsilon(3S)$
on-resonance
sample, we count the number of on-resonance hadronic events
and subtract the number of off-resonance events scaled by
the ratios of luminosities and cross sections between the
on- and off-resonance samples. The luminosity ratio is determined
from diphoton events. 
The number of $\Upsilon$ mesons is~\cite{McGregor:2008ek}
\beq
N_{\Upsilon} = \left( N_{\rm had} - \kappa  \, N_{\rm had}^\off 
   {N_{\gaga} \over  N_{\gaga}^\off}
              \right) 
   {1 \over \epsilon_{\rm had}},
\label{eq:nupsilon}
\eeq
where $N_{\rm had}$ ($N_{\rm had}^\off$) is the number of events
satisfying the $\epem\to{\rm hadrons}$ selection criteria in the
on-resonance (off-resonance) sample, 
$N_{\gaga}$ ($N_{\gaga}^\off$) is the number of events
satisfying the $\epem\to\gaga$ selection criteria in the
on-resonance (off-resonance) sample, 
$\epsilon_{\rm had}$ is the reconstruction efficiency for the on-resonance
hadronic events, and $\kappa$ is a correction factor accounting for the 
small $s$-dependence of the visible cross sections of the continuum 
hadronic and $\gaga$ events.

Using Eq.~(\ref{eq:nUps-to-ee}), we determine that
$\Upsilon\to\epem X$ background constitutes $(1.4\pm 0.1)\%$ of
the events passing the $\eeMode$ selection 
in the on-resonance $\Upsilon(2S)$ sample and
$(0.9\pm 0.1)\%$ in the $\Upsilon(3S)$ sample. 
The uncertainties are dominated by the uncertainties on the $\Upsilon\to\epem X$
branching fractions. 
The uncertainty on $N_{\Upsilon}$
is 0.9\%, dominated by the determination of $\epsilon_{\rm had}$,
and has a negligible effect on the $N_{\Upsilon\to\epem X}$ uncertainty.

In the dimuon channel,  $\Upsilon\to\mu^+\mu^-$ events constitute $(21.9 \pm
2.2)\%$ of the selected $\mumuMode$ candidate events for the $\Upsilon(2S)$
sample and $(14.3 \pm 1.4)\%$ for the $\Upsilon(3S)$ sample.  Due to
the large uncertainty introduced by this background, dimuon events are
not used for Run~7, as mentioned above.

\subsection{Visible Cross Sections}
\label{sec:sigma}

The visible cross sections $\sigmavis$ (see Eq.~(\ref{eq:sigmavis}))
for Bhabha and dimuon events are initially obtained from the MC
simulation for each run period and CM energy\footnote{The MC
generators are not valid in some parts of phase space, in particular
for small-angle Bhabha scattering, which is excluded by the analysis 
selection criteria. Therefore, the simulation can
be used to evaluate the visible cross section, but not the full cross
section and efficiency separately.}.
We then correct the values of $\sigmavis$ for small data-MC efficiency differences,
determined as follows.

We determine the inefficiency  
of the trigger and offline-filter selection from the fraction of
events that fail this selection but satisfy the final selection
requirements, using event samples that are allowed to bypass the
level-3 trigger and offline filter. From the inefficiency difference between data
and MC, we apply run-by-run corrections to $\sigmavis$ of up to 0.3\%.

The track-reconstruction inefficiency is measured from the fraction of 
Bhabha events in which only one track is found. To minimize the non-Bhabha
events in this sample, one of the tracks must satisfy tight selection criteria:
$0.95 < P < 1.05$,
$0.9 < (E/p) < 1.1$, and 
$\left|\cos\theta\right|<0.70\rad$. 
A second track is not found in 0.2\% of these events.  The
identification of these one-track events as \eeMode\ is justified by
the observation that the highest-energy EMC cluster, 
other than the cluster associated with the track, has CM
acolinearity with respect to the track of no more than about
$10^\circ$ (some acolinearity is expected, since the missed track bends in 
the magnetic field), 
and that the ratio between the energy of this cluster to
the track momentum peaks at 1.  From the data-MC inefficiency
difference, we apply run-dependent corrections to $\sigmavis$ in the
range 0.14\%-0.27\%.

Table~\ref{tab:xsections} shows the corrected visible cross sections
for the different \pep2\ CM energies. For Runs~1--6, we observe a
run-to-run variation of $\pm 0.21\%$ ($\pm 0.7\%$) in the value of
\sigmavis\ for the Bhabha (dimuon) channel.

%
%
\begin{table*}[htbp]
\caption{\label{tab:xsections} Visible cross section $\sigmavis$ 
(see Eq.~(\ref{eq:sigmavis})),
with the relative uncertainty in percent shown in parentheses,
for the different data-taking periods
categorized according to the 
center-of-mass energy $\sqrt{s}$, which was equal to (``On'') or just below (``Off'') 
the masses of the $\Upsilon$ resonances. Results for the $\Upsilon(4S)$
samples are luminosity-averaged over Runs~1--6.
The uncertainties are systematic and are described in
Section~\ref{sec:syst}.  }  
\vspace{4mm}
\centering
   \begin{tabular}{c|c|c}
     \hline\hline
Sample & \multicolumn{2}{c}{$\sigmavis$ (nb)} \cr
                    \cline{2-3}
           &    \eeMode           &         \mumuMode    \cr
     \hline
 On $\Upsilon(4S)$  & $6.169 \pm 0.041$ ~~~(0.7) & $0.4294 \pm 0.0023$ ~~~(0.5) \cr
 Off $\Upsilon(4S)$ & $6.232 \pm 0.044$ ~~~(0.7) & $0.4333 \pm 0.0025$ ~~~(0.6) \cr
 On $\Upsilon(3S)$  & $6.461 \pm 0.037$ ~~~(0.6) & $0.4488 \pm 0.0028$ ~~~(0.6) \cr
 Off $\Upsilon(3S)$ & $6.508 \pm 0.056$ ~~~(0.9) & $0.4501 \pm 0.0040$ ~~~(0.9) \cr
 On $\Upsilon(2S)$  & $6.933 \pm 0.042$ ~~~(0.6) & $0.4802 \pm 0.0030$ ~~~(0.6) \cr
 Off $\Upsilon(2S)$ & $6.866 \pm 0.051$ ~~~(0.7) & $0.4721 \pm 0.0036$ ~~~(0.8) \cr
    \hline\hline
   \end{tabular}
\end{table*}

\section{Systematic Uncertainties}
\label{sec:syst}

Table~\ref{tab:syst} summarizes the systematic uncertainties, which
are described in detail below.

\begin{table*}
\caption{\label{tab:syst} Relative systematic uncertainties on the measured 
integrated luminosity. }
\vspace{4mm}
\centering
\begin{tabular}{l|l}
     \hline\hline
Source & Relative uncertainty on \lumi\ (\%)  \cr
\hline
Theoretical cross section  &   0.26 ($\epem$), 0.44 ($\mu^+\mu^-$) \cr
Track-reconstruction efficiency &  0.13 (Runs~1--6), 0.20 (Run~7) \cr
Trigger \& offline-filter efficiency  &  0.10   \cr
Data-MC differences & 0.5--0.7 ($\epem$), 0.24--0.28 ($\mu^+\mu^-$) \cr
Time dependence &  0.16--0.46 (Off-resonance) \cr
Background subtraction & 0.02 (Runs~1--6), 
                   0.10 ($\Upsilon(3S)$), 0.15 ($\Upsilon(2S)$) \cr
Boost uncertainty &  0.2 (Run~7) \cr
     \hline\hline
\end{tabular}
\end{table*}

For the selection criteria used in this analysis, we find that the
cross section reported by BHWIDE is consistent with that of the
BABAYAGA~\cite{ref:babayaga} generator to within the statistical
uncertainty of the comparison, $0.06\%$.  We add this uncertainty in
quadrature to the BABAYAGA theoretical uncertainty of
0.20\%~\cite{ref:babayaga} to obtain the total uncertainty of 0.21\%.
The uncertainty on the dimuon cross sections is taken to be 0.44\%,
based on Ref.~\cite{Banerjee:2007is}.

From the data-MC comparisons described in
Section~\ref{sec:sigma},
we estimate an uncertainty of 0.13\% (0.20\%)  
for the track-reconstruction
efficiency for Runs~1--6 (Run~7), 
corresponding to approximately half the largest correction within these data samples.  
An uncertainty of 0.1\% is estimated for the trigger and offline-filter efficiency
correction 
by rounding up the largest of the run-dependent statistical uncertainties
of this correction.
To account for differences between the distributions of data and MC
events in the variables used for event selection, we vary the
selection requirements over wide ranges throughout the tails of the
signal-event distributions, and repeat the full analysis for each
variation.  For each selection variable, the largest resulting change
in \lumi\ is taken to be the associated uncertainty. The uncertainties
for the different selection variables are added in quadrature for each
run, with resulting uncertainties ranging between $0.5\%$ and $0.7\%$
for $\eeMode$ and between $0.24\%$ and $0.28\%$ for $\mumuMode$.

\begin{figure}[htp!]
  \includegraphics[width=\linewidth]{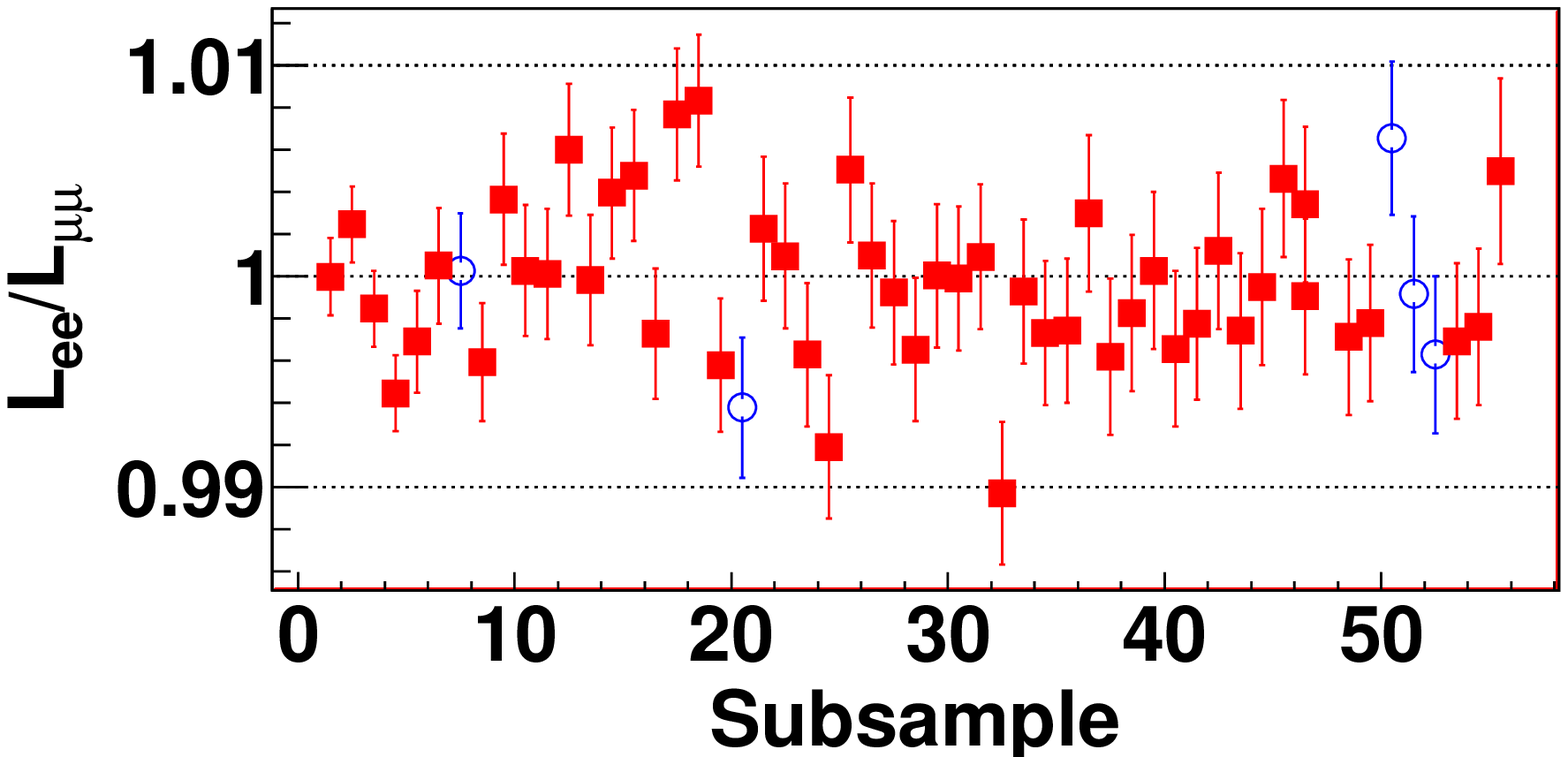} \\
  \includegraphics[width=\linewidth]{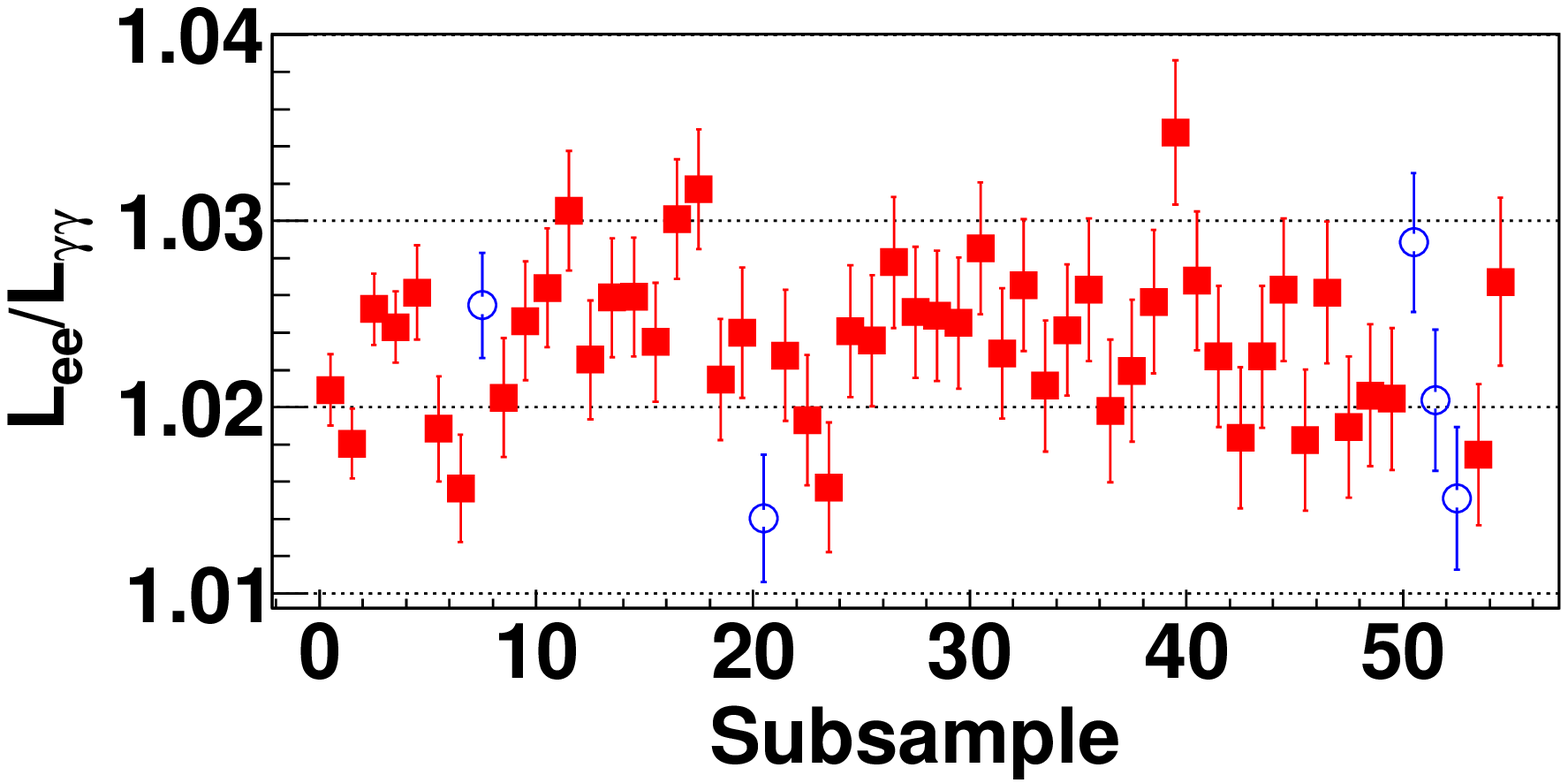}
\caption{\label{figtimedep} (Top) The ratio $\lumi^{ee} / \lumi^{\mumu}$
of the luminosity values computed with Bhabha and dimuon events and
(bottom) the ratio $\lumi^{ee} / \lumi^{\gaga}$ obtained with Bhabha
and diphoton events for subsamples of Run~4. Off-resonance (on-resonance) 
subsamples are indicated with blue open circles (red filled squares).
The luminosity ratios
shown do not include the efficiency corrections discussed in the
text.}
\end{figure}

The luminosity and systematic uncertainties are evaluated for the
entire period of data collection for each particular run.  Use of
subsamples within a run may introduce time-dependent variations in
efficiency that are not accounted for in the analysis.  In particular,
off-resonance data are collected at relatively rare intervals, and
could therefore be subject to such time-dependent effects. Therefore,
we estimate an additional systematic uncertainty for the off-resonance
luminosity, accounting for tracking-related and EMC-related time
variation studied using the on-resonance samples.
The on-resonance data sample for each run is divided into at least ten
subsamples with luminosities of about
1 to $2~\invfb$ each. In each subsample $i$, we calculate the ratio
$x_i = \lumi_i^{ee} / \lumi_i^{\mu\mu}$ of the luminosity values
obtained with Bhabha and dimuon events.
We use the spread in the $x_i$ values, after subtraction of the
estimated statistical component of the spread, to estimate the
off-resonance luminosity uncertainty associated with the time
variation of any EMC-related effects.
Similarly, we use the spread of the ratios $\lumi_i^{ee} /
\lumi_i^{\gaga}$ of the luminosity values obtained with Bhabha and
diphoton events to estimate the uncertainty due to the time variation
of tracking-related effects. Finally, these two uncertainties are added
in quadrature.
As an illustration, the values of $\lumi_i^{ee} / \lumi_i^{\mumu}$ and
$\lumi_i^{ee} / \lumi_i^{\gaga}$ for the different subsamples of Run~4
are shown in Fig.~\ref{figtimedep}.

The uncertainties on the background subtraction, described in
Section~\ref{sec:bgd}, are propagated to the final uncertainty on
\lumi.  For Run~7, we estimate an additional uncertainty of 0.2\% on
the signal reconstruction efficiency, arising from the uncertainty on the
laboratory-to-CM boost associated with changing the \pep2\
energy from the $\Upsilon(4S)$ to the $\Upsilon(3S)$ and 
$\Upsilon(2S)$.

Systematic uncertainties from the different sources are added in quadrature,
separately for each channel (\eeMode, \mumuMode), run, and on/off-resonance 
data-taking period. 
When combining results in Section~\ref{sec:results}, we 
take into account the following correlations between systematic uncertainties.
The uncertainties on the track-reconstruction efficiency 
and on the trigger and offline-filter efficiency are positively correlated between 
the two channels.
Uncertainties in the theoretical cross section, background
subtraction, trigger and offline-filter efficiencies, and selection-criteria
variation are positively correlated for the different runs, as well as for the
on-resonance and off-resonance periods.

\section{Results}
\label{sec:results}

Table~\ref{tab:lumi-summary} lists the integrated luminosity results
for the on- and off-resonance samples for each run. The results for
Runs~1--6 are averaged over the \mumu\ and $\epem$ channels,
accounting for correlated uncertainties.  The results obtained with
the two modes are compatible and have similar overall uncertainties,
with the \mumu\ uncertainties being somewhat smaller.  (As noted in
Section~\ref{sec:analysys-method}, the Run-7 luminosity is obtained
with $\epem$ events only.) The ratios between the on-resonance and
off-resonance integrated luminosities are also given.
Table~\ref{tab:lumi-1--6-detail} shows a run-by-run breakdown of the
results for the $\Upsilon(4S)$ periods.

\begin{table*}[hbt]
\caption{\label{tab:lumi-summary} The integrated luminosities of 
the on-resonance ($\lumi_{\rm on}$)
and off-resonance ($\lumi_{\rm off}$) data samples recorded at and just 
below the $\Upsilon$
resonances, and the ratio between the on- and off-resonance 
integrated luminosities. 
For each entry, the first uncertainty is statistical, the second uncertainty is systematic,
and the total relative uncertainty in percent is given in parentheses.}
 \vspace{4mm}
\centering
\begin{tabular}{c|c|c|c}
\hline\hline 
Resonance    & {$\lumi_{\rm on}$~(\invfb)} & {$\lumi_{\rm off}$~(\invfb)} & {$\lumi_{\rm on}/\lumi_{\rm off}$} \\
\hline      
$\Upsilon(4S)$ & $424.18 \pm 0.04 \pm 1.82$ ~~~(0.43) & $43.92 \pm 0.01 \pm 0.19$   ~~~(0.43) & $9.658 \pm 0.003 \pm 0.007$ ~~~(0.08) \\
$\Upsilon(3S)$ & $27.96 \pm 0.03 \pm 0.16$  ~~~(0.58)  & $2.623 \pm 0.008 \pm 0.017$ ~~~(0.72)  & $10.66 \pm 0.03 \pm 0.03$   ~~~(0.40) \\
$\Upsilon(2S)$ & $13.60 \pm 0.02 \pm 0.09$  ~~~(0.68)  & $1.419 \pm 0.006 \pm 0.011$ ~~~(0.88)  & $9.58 \pm 0.04 \pm 0.04$    ~~~(0.59)  \\
\hline\hline 
\end{tabular}
\end{table*}

\begin{table*}[hbt]
\caption{\label{tab:lumi-1--6-detail} The on-resonance ($\lumi_{\rm on}$)
and off-resonance ($\lumi_{\rm off}$) integrated luminosities of the individual $\Upsilon(4S)$ runs,
and the ratio between the on- and off-resonance integrated luminosities. 
For each entry, the first uncertainty is statistical, the second uncertainty is systematic,
and the total relative uncertainty in percent is given in parentheses.}
\vspace{4mm}
\centering
\begin{tabular}{c|c|c|c}
\hline \hline 
Run     &  {$\lumi_{\rm on}$~(\invfb)} & {$\lumi_{\rm off}$~(\invfb)} & {$\lumi_{\rm on}/\lumi_{\rm off}$} \\
\hline 
1  &  $20.37 \pm 0.01 \pm 0.09$  ~~~(0.44) & $2.564 \pm 0.002 \pm 0.014$   ~~~(0.55)  &  $7.946 \pm 0.006 \pm  0.027$ ~~~(0.35)\\
2  &  $61.32 \pm 0.01 \pm 0.26$  ~~~(0.42) & $6.868 \pm 0.004 \pm 0.034$   ~~~(0.44) &  $8.928 \pm 0.006 \pm  0.023$ ~~~(0.27)\\
3  &  $32.28 \pm 0.01 \pm 0.13$  ~~~(0.40) & $2.443 \pm 0.003\pm 0.012$    ~~~(0.51)  &  $13.213 \pm 0.015 \pm 0.037$ ~~~(0.30)\\
4  &  $99.58 \pm 0.02 \pm 0.41$  ~~~(0.41) & $10.016 \pm 0.007 \pm 0.043$  ~~~(0.43) &  $9.943 \pm 0.007 \pm 0.012$  ~~~(0.14)\\
5  &  $132.33 \pm 0.02 \pm 0.59$ ~~~(0.45) & $14.278 \pm  0.008 \pm 0.066$ ~~~(0.47)  & $9.268 \pm 0.005 \pm 0.012$   ~~~(0.14)\\
6  &  $78.31 \pm 0.02 \pm 0.35$  ~~~(0.45) & $7.752 \pm 0.006 \pm 0.036$   ~~~(0.47)  & $10.102 \pm  0.008 \pm 0.013$ ~~~(0.15)\\
\hline \hline 
\end{tabular}
\end{table*}

\section{Acknowledgements}
We are grateful for the 
extraordinary contributions of our \pep2\ colleagues in
achieving the excellent luminosity and machine conditions
that have made this work possible.
The success of this project also relies critically on the 
expertise and dedication of the computing organizations that 
support \babar.
The collaborating institutions wish to thank 
SLAC for its support and the kind hospitality extended to them. 
This work is supported by the
US Department of Energy
and National Science Foundation, the
Natural Sciences and Engineering Research Council (Canada),
the Commissariat \`a l'Energie Atomique and
Institut National de Physique Nucl\'eaire et de Physique des Particules
(France), the
Bundesministerium f\"ur Bildung und Forschung and
Deutsche Forschungsgemeinschaft
(Germany), the
Istituto Nazionale di Fisica Nucleare (Italy),
the Foundation for Fundamental Research on Matter (The Netherlands),
the Research Council of Norway, the
Ministry of Education and Science of the Russian Federation, 
Ministerio de Ciencia e Innovaci\'on (Spain), and the
Science and Technology Facilities Council (United Kingdom).
Individuals have received support from 
the Marie-Curie IEF program (European Union) and the A. P. Sloan Foundation (USA).


\end{doublespace} 
\end{document}